\documentclass{aa}
\usepackage{psfig,longtable,lscape}
\newcommand{\et}{et al.}
\newcommand{\ct}{cts s$^{-1}$}
\newcommand{\perr}{r$_{90}$}

\newcommand{\ext}{r$_{\rm ext}$}

\newcommand{\extl}{ML$_{\rm ext}$}
\newcommand{\exil}{ML$_{\rm exi}$}

\begin{document}

\thesaurus{
           04.03.1; 
           11.13.1; 
           11.19.5; 
           13.25.2; 
           13.25.5  
          }

\title{ROSAT HRI catalogue of X-ray sources in the SMC region}

\author{Manami Sasaki, Frank Haberl, and Wolfgang Pietsch}
\authorrunning{Sasaki et al.}

\offprints{M.\ Sasaki (manami@mpe.mpg.de)}

\institute{Max--Planck--Institut f\"ur extraterrestrische Physik,
 Giessenbachstra{\ss}e, 85748 Garching, Germany}

\date{Received March 24, 2000; accepted August 25, 2000}

\maketitle

\begin{abstract}
During the operational phase of the ROSAT satellite between 1990 and
1998 the X-ray telescope pointed 71 times to the Small Magellanic
Cloud (SMC) for observations with the High Resolution Imager
(HRI), covering a field of 5\degr\ x 5\degr\,. From this data a
catalogue of 121 discrete X-ray sources was derived. By 
cross-correlating the source catalogue with the SIMBAD data base and
the TYCHO catalogue, the systematic positional error of the HRI source
positions could be reduced. In total the X-ray position for 99 HRI
sources was corrected yielding positional errors between 1\arcsec\
and 16\arcsec\,.   
The HRI catalogue was also cross-correlated with the catalogues derived
from the ROSAT Position Sensitive Propotional Counter (PSPC) pointings
(Kahabka \et\ 1999, Haberl \et\ 2000). For 75 HRI sources PSPC
counterparts were found and thus hardness ratios are given.   
With the help of the information obtained from the
cross-correlations 56 HRI sources were identified with 
objects of known or proposed nature. 
Four foreground stars, six supernova remnants, four supersoft sources, 
12 X-ray binaries, and one AGN were detected by the HRI. 
Based on the existence of a likely optical counterpart 
or properties like hardness ratio and X-ray to optical flux ratio, 
further 15 HRI sources were classified into different source types.

\keywords{Catalogues -- Galaxies: Magellanic Clouds --
          Galaxies: stellar content --
          X-rays: galaxies -- X-rays: stars}
\end{abstract}

\section{Introduction}

Since the Magellanic Clouds (MCs) are among the closest galaxies to
the Milky Way, we are able to study their overall characteristics like
source distribution as well as properties of single objects within the
galaxies. Complementary to the Large Magellanic Cloud (LMC)
source catalogue of the ROSAT High Resolution Imager (HRI)
published by Sasaki \et\ (hereafter SHP00) HRI data of the
Small Magellanic Cloud (SMC) region is analyzed in the work presented
here. The SMC with a distance of $\sim$60\,kpc has been subject to
many studies in different wavelength bands (e.g.\ Davies \et\ 1976,
Kennicutt \& Hodge 1986, Staveley-Smith \et\ 1997).

X-ray emission from the MCs was first observed in
a rocket mission in 1970 (Price et al.\ 1971). The first discovered
single X-ray source in the Small Magellanic Cloud (SMC) was SMC X-1
detected in an Uhuru observation (Leong et al.\ 1971). 
The Einstein Observatory performed several X-ray observations of the
SMC making a thorough survey (Seward \& Mitchell 1981, Inoue et al.\
1983, Bruhweiler et al.\ 1987, Wang \& Wu 1992). Pointed
observations by the X-ray satellite ROSAT (Tr\"umper 1982) covered the
MCs in the energy range of 0.1 -- 2.4 keV. Kahabka \et\
(1999, hereafter KPFH99) presented the first ROSAT PSPC source
catalogue of the SMC. After analyzing all the available PSPC pointings
Haberl \et\ (2000, hereafter HFPK00) published a ROSAT PSPC
catalogue of discrete sources in the SMC containing 517 X-ray sources.

\section{ROSAT HRI data}

\subsection{Source detection}

\begin{figure}[t]
\centerline{\psfig{figure=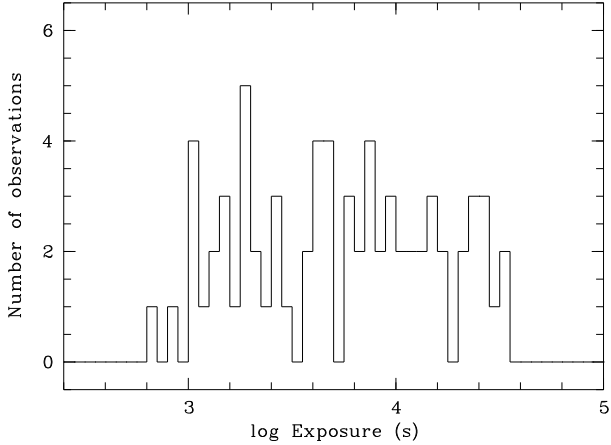,angle=270,width=9cm}}
\caption[]{\label{exphisto} Histogram for HRI pointing exposure times
of SMC observations.}
\end{figure}

In a 5\degr\ x 5\degr\ field around RA = 01$^h$ 00$^m$ 00$^s$, Dec =
$-73\degr\ 00\arcmin\ 00\arcsec$ (J2000.0) including the whole SMC 
a total of 71 pointings were performed by the ROSAT HRI (David \et\
1996) between 1991 and 1998. The total exposure time is shown in Fig.\
\ref{exposure} as 
contours plotted over a grey scale PSPC image (HFPK00). The number of
the analyzed HRI pointings with integration times higher than 100
seconds is demonstrated in Fig.\ \ref{exphisto}. 

\begin{figure*}[t]
\caption[]{\label{exposure} Contours of total exposure times of the
HRI pointings plotted on a grey scale PSPC image (0.1 -- 2.4 keV). The
contours signify regions of exposure times with 1, 20, 40, 60, 80,
and 100 ksec.} 
\end{figure*}

For data analysis and correction of X-ray positions same procedures as
developed for the LMC (see SHP00) were applied to the SMC data using
the EXSAS software package (Zimmermann \et\ 1994). 
For each pointing sources were detected in two sliding window methods
using the local background and a spline fitted background map. Both
detection lists served as input for the following maximum
likelihood method. Detections with existence likelihood higher than 10.0
and telescope off-axis angle smaller than 15\arcmin\
were accepted as X-ray sources. 
In addition pointings with directions within a radius of 1\arcmin\ were
merged in order to increase the significance of source signals. For 18
regions in the SMC new deeper data could be created. Further faint
sources were found by the source detection procedure in the merged
data and augmented the number of detected HRI sources.
Finally the source lists from the single and coadded pointings were
merged. Multiple detections of one source in different pointings
were reduced to one catalogue entry choosing the detection with the highest
positional accuracy. In the end we obtained a HRI source catalogue with 
121 sources in the SMC region. 

This source catalogue was cross-correlated with the PSPC catalogue
(HFPK00). We looked for coincidences with objects in the
SIMBAD data base operated at the Centre de Donn\'ees astronomiques de
Strasbourg, the TYCHO catalogue from the ESA Hipparcos space
astrometry satellite (Hoeg \et\ 1997), 
and the point source catalogue from the Deep Near Infrared Survey of the 
Southern Sky (DENIS, Cioni \et\ 2000, hereafter CLH00) 
in order to identify X-ray sources. The sources were also compared to 
objects known from
literature like those published by KPFH99 or Schmidtke \et\ (1999). 
Table \ref{wholecat} lists the HRI sources with corrected coordinates
(see Sec.\ \ref{position}), positional error, existence likelihood,
HRI count rate, extent, extent likelihood, the corresponding PSPC
source with count rate and hardness ratios (HFPK00). 

\subsection{Source position}\label{position}

For ROSAT the systematic positional error (boresight error) is about
7\arcsec\,. In cases where an X-ray source in a
pointed observation can be identified with an object with accurately
known coordinates, the systematic error can be reduced for all
detected X-ray sources from the same observation. The remaining
systematic error for all sources in that pointing is
\begin{equation}
\sigma_{sys,p} = \sqrt{\sigma_{opt}^2+\sigma_{stat,s}^2}\ ,
\end{equation}
$\sigma_{opt}$ being the error of the optical position and
$\sigma_{stat,s}$ the statistical 90\% confidence error of the
identified X-ray
source used for correction. A systematic error of 7\arcsec\ was used
for all other sources of which the X-ray position could not be
corrected. The total positional error is finally given by:
\begin{equation}
\sigma_{tot,X} = \sqrt{\sigma_{sys,p}^2+\sigma_{stat,X}^2}\ .
\end{equation}

Therefore the positions of HRI sources identified with TYCHO catalogue
sources were corrected to more accurate optical coordinates. For X-ray
sources already discussed in the literature and for which the
optical counterpart is known, the X-ray positions were also
corrected. For each correction the identification of the HRI source
with an optical object was confirmed on Digitized Sky Survey 
(DSS2, red) frames.  
With the help of the corrected X-ray source positions the coordinates
of all the pointings observing these sources were newly computed,
and for other sources detected in the corrected pointings again more
accurate coordinates were determined. In doing so the X-ray positions
of 99 out of 121 sources were improved.

Before the corrections the mean positional error of the 121 HRI
catalogue sources was 8\farcs1.		
After the boresight correction the mean positional error was reduced
to 5\farcs6, and 4\farcs9 for corrected sources only.

\subsection{Flux variability}\label{variability}

\begin{table*}[t]
\caption[]{HRI sources with significant flux variability}
\begin{tabular}{rrrrrrrl}
\hline\noalign{\smallskip}
\multicolumn{1}{c}{1} & \multicolumn{1}{c}{2} & \multicolumn{1}{c}{3} & \multicolumn{1}{c}{4} &
\multicolumn{1}{c}{5} & \multicolumn{1}{c}{6} & \multicolumn{1}{c}{7} & ~~8 \\
\hline\noalign{\smallskip}  
\multicolumn{1}{c}{No} & \multicolumn{1}{c}{Rate} &
\multicolumn{1}{c}{Rate} & 
\multicolumn{1}{c}{$\frac{\rm F_{max}}{\rm F_{min}}$} &
\multicolumn{1}{c}{Red.\ $\chi^2$} & \multicolumn{1}{c}{DOF} & 
\multicolumn{1}{c}{No} & Remarks \\
 & \multicolumn{1}{c}{HRI} & \multicolumn{1}{c}{PSPC} & &
 & & \multicolumn{1}{c}{PSPC} & \\
 & \multicolumn{1}{c}{[\ct]} & \multicolumn{1}{c}{[\ct]} & &
 & & & \\
\noalign{\smallskip}\hline\noalign{\smallskip} 
  7 & 1.0e-01 & 4.4e-01 &  2.5 &   112.1 &       8 &  176 & SSS 1E0035.4-7230                                                                \\ 
 10 & 4.4e-03 & 1.6e-02 &  2.0 &     8.1 &       2 &  562 & foreground star HD 3880                                                          \\ 
 23 & 1.7e-02 & 7.2e-03 &  2.9e+01 &   475.1 &       3 &  512 & SSS RXJ0048.4-7332                                                               \\ 
 35 & 1.0e-02 &        &   3.7e+02 &    14.1 &       3 &     & $<$foreground star$>$ SkKM 62                                             \\ 
 44 & 4.9e-03 & 2.4e-02 &  5.5e+03 &  2973.0 &       6 &  453 & HMXB Be/X RXJ0052.1-7319, pulsar [LPM99],[ISC99]                                 \\ 
 46 & 3.0e-03 & 1.5e-02 &   4.5e+02 &     9.4 &       9 &   94 & HMXB Be/X RXJ0052.9-7158 [CSM97],[SCC99],[HS00]                                    \\ 
 51 & 2.7e-03 &        &   3.1e+02 &    95.0 &       4 &  242 & HMXB? XTEJ0053-724, pulsar [CML98]                                               \\ 
 65 & 4.3e-03 & 1.9e-02 &  2.4 &    10.4 &       3 &  508 & \\
 74 & 2.2e-03 & 7.9e-03 & 2.5e+04 &     8.7 &       9 &  136 & HMXB? Be/X [HS00]                                                           \\ 
 79 & 9.4e-02 & 3.6e-01 &  2.6 &   115.5 &       5 &   47 & SSS 1E0056.8-7146                                                                \\ 
 93 & 2.9e-04 & $<$9.1e-03 &   3.5e+02 &     9.9 &       7 &  132 & HMXB Be/X RXJ0101.0-7206 [KP96],[SCB99]                                          \\ 
 95 & 2.9e-03 & 2.4e-02 &  1.2e+01 &     8.1 &       9 &  159 & HMXB? Be/X [HS00]                                                           \\ 
101 & 4.3e-03 & 1.6e-02 &  2.2 &     8.3 &       3 &  143 & HMXB SAXJ0103.2-7209, pulsar [ISC98],[HS94]                          \\ 
114 & 1.6e-02 &        &   6.7e+01 &   119.6 &       4 &     &                                                                                  \\ 
118 & 3.7e-01 & 4.9e+00 &   2.1e+02 & 24762.9 &       8 &  482 & HMXB SMC X-1                                                                     \\ 
120 & 3.4e-02 & 1.1e-01 &  2.6 &    12.1 &       8 &  478 & foreground star HD 8191 [CSM97]                                                  \\ 
\noalign{\smallskip}
\hline
\end{tabular}
\label{variable}

Notes to columns No 2 and 3:
Count rates are the mean of output values from maximum likelihood
algorithm for single pointings. 
Count rate with $<$ is mean upper limit.

Notes to column No 6:
Degrees of freedom.

Notes to column No 7:
Source number from HFPK00.

Notes to column No 8:
Sources classified in this work are put in $<$ $>$.
Abbreviations for references in square brackets are given in
the literature list.
\end{table*}

\begin{figure}
\centerline{\psfig{figure=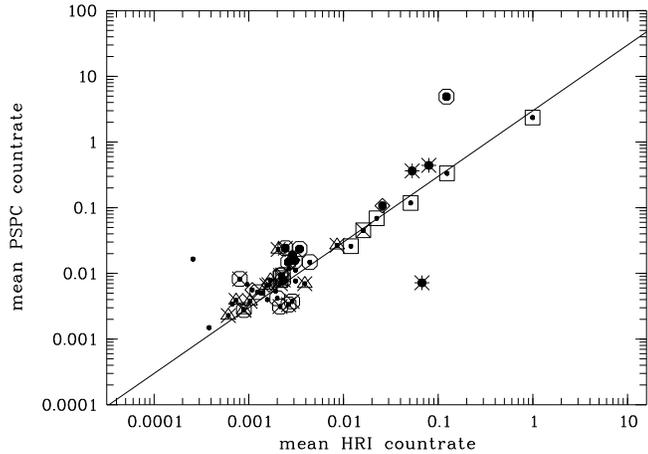,angle=270,width=9cm}}
\caption[]{\label{rate} The mean of observation-averaged count rates
from PSPC pointings versus HRI count rates: 
All sources observed by both detectors are plotted with small dots. In 
addition SNRs are marked with squares, XBs with circles, SSSs with 
asterisks, AGN with triangles, and stars with lozenges.
Crossed symbols indicate already known candidates and variable sources
are additionally marked with larger dots. 
The line is drawn for a PSPC/HRI conversion factor of 3.}
\end{figure}

In the maximum likelihood algorithm for source detection, count rates
are computed for point sources with intensity
distribution peaking in the center. For extended sources the values
resulting from the maximum likelihood algorithm are not reliable,
because the count rates are determined by fitting a Gaussian to the
source intensity profile. Thus for extended sources the total count rate
was interactively integrated within a circle around the source, 
subtracting the background from a ring around the source.
For faint sources upper limits were calculated by the maximum
likelihood algorithm.

109 HRI sources were detected in more than one ROSAT observation and
allowed flux variability studies on time scales of days to years.
Lightcurves for these HRI sources were produced
in order to test the flux variability. For this purpose not
only HRI pointings were used, but also the PSPC pointings of the SMC
region. From PSPC count rates and upper limits corresponding HRI values
were computed by dividing the PSPC values by 3 which is a typical
conversion factor for the ROSAT detectors (see SHP00). Figure \ref{rate}
shows the correlation between the ROSAT detector count rates
for sources detected both by the HRI and the PSPC.

A $\chi^2$-test and the ratio between maximum and minimum flux give
information on the significance of variability in the lightcurves
(HP99a). Variations with reduced $\chi^2$ higher than 5.0 were
accepted as significant. Finally the list of variable sources contains
16 HRI sources, including three supersoft sources (SSSs) and eight
X-ray binaries (XBs) and candidates. Table \ref{variable} lists
parameters indicating variability and the mean count rates both for
HRI and for PSPC observations for variable HRI sources. 

\section{Source classes}

Various source classes with different origin of the X-ray emission are
found in the MC regions:
Foreground stars in the Galaxy, supernova remnants (SNRs), SSSs,
XBs, and background objects like galaxies and clusters. 
The X-ray sources detected by the
HRI can be assigned to the aforementioned classes if they are
identified with known objects from literature (Sec.\
\ref{identifiedsou}) or if their distinguished X-ray properties allow
this. Seventy five out of 121 HRI catalogue sources were 
also detected by the PSPC. We studied the X-ray properties of the
identified sources to obtain tools for classifying detections by
the HRI (Sec.\ \ref{newclass}). 

\subsection{Source identification}\label{identifiedsou}

\begin{figure}[t]
\centerline{\psfig{figure=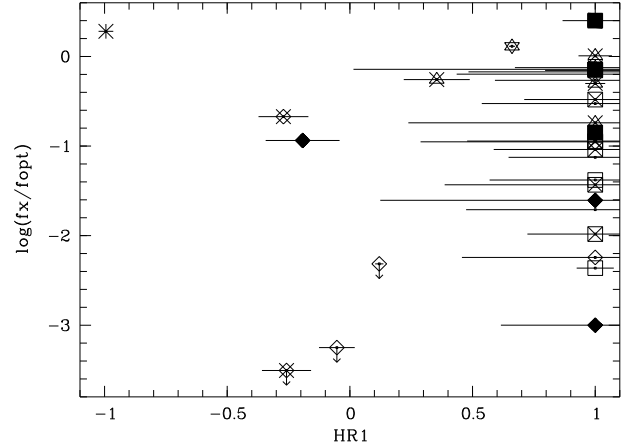,angle=270,width=9cm}}
\caption[]{\label{logfxfopt} Flux ratio log(f$_{x}$/f$_{opt}$) over 
hardness ratio 1: Asterisks indicate SSSs, stars SNRs, open
squares XBs, open lozenges stars, and open triangles AGN,
which are all known objects from literature. 
Already known candidates are marked with crossed symbols and new
classifications with filled symbols.} 
\end{figure}

56 HRI sources were identified with known SMC objects, foreground
stars, or AGN based on their accurate positions from HRI
observations and the X-ray properties like extent or PSPC hardness
ratios (HFPK00) as it is shown in Table \ref{identified}. 
The accurate HRI positions make it possible to identify likely
optical counterparts. For optical sources within the error circle
of the X-ray position, the B and R magnitudes were determined from the
USNO-A2.0 catalogue produced by the United States Naval Observatory
(Monet 1998). 
The flux ratio was computed from the B magnitude
and the X-ray count rates applying the equation 
\begin{eqnarray}
log({\rm f}_{x}/{\rm f}_{opt}) & = & log(3\cdot{\rm HRI\ counts/s}\cdot
10^{-11}) \nonumber \\
 & & +0.4B+5.37 
\end{eqnarray}
(Maccacaro et al.\ 1988; HP99b; SHP00). 
Originally this value was calculated for V magnitude. But since the V 
magnitude could only be determined for a small sample, we decided to use
the B magnitude. It gives a clue for identifying foreground stars, because 
for stars log(f$_{x}$/f$_{opt}$) usually is negative. For very bright 
stars the B magnitude is not given in the USNO-A2.0 
catalogue. In those cases {\it B} was set to 11.0 for calculating 
log(f$_{x}$/f$_{opt}$) yielding an upper limit.
In Fig.\ \ref{logfxfopt} the log(f$_{x}$/f$_{opt}$) values are plotted over 
the PSPC hardness ratio 1 for HRI sources with likely optical counterpart 
and also detected by the PSPC.

\subsubsection{Foreground stars}

In addition to four foreground stars, two known candidates (sources No
18 and 86) were detected by the HRI. No 18 and 86 were detected
by the PSPC and classified as foreground stars by KPFH99 and FHP00.  
The HRI observations give positions more accurate than the PSPC (total
positional error of 5\farcs6 and 7\farcs2, respectively). At the
HRI positions an optical source was found for each 
case, with {\it R} = 11.3, log(f$_{x}$/f$_{opt}$) = --3.51 for No 18
and {\it R} = 13.1, log(f$_{x}$/f$_{opt}$) = --0.67 for No 86
confirming the proposed classification.

\subsubsection{Supernova Remnants}

\begin{figure}[t]
\centerline{\psfig{figure=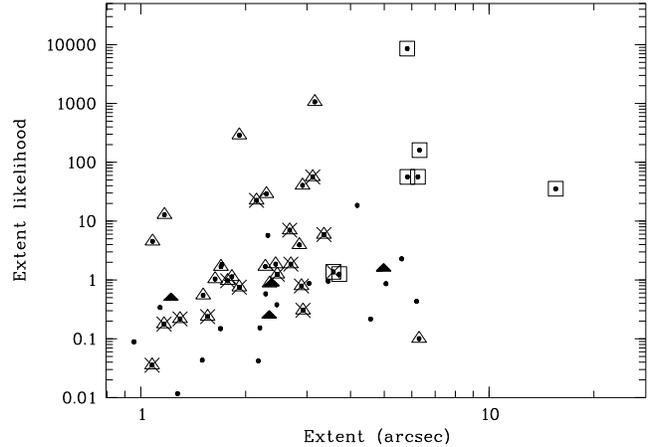,angle=270,width=9cm}}
\caption[]{\label{extent} Source extent is plotted over extent likelihood 
as small dots for all SMC sources, for which the HRI measured an extent 
higher than zero. SNRs are additionally marked with squares 
and known point sources with triangles. 
Open symbols are used for identified sources,
crossed symbols for candidates known from literature, 
and filled symbols for new classifications.}
\end{figure}

Six SNRs were detected by the HRI. For these SNRs extents larger
than 5\farcs8 were measured by the HRI with a likelihood higher
than 35 (see Fig.\ \ref{extent}), except for SNR 0056-72.5 which has no
significant extent. 

Source No 22 is identified with a PSPC source which was suggested as 
SNR candidate by KPFH99. The extent of the HRI source calculated in
the maximum likelihood algorithm is only 3\farcs6. However, the HRI
image shows that the source is only the brightest knot in an extended
emission, confirming the classification as a SNR candidate.

\subsubsection{Supersoft sources}

Four SSSs were found in the HRI observations which were also detected in PSPC
observations.
For three of the sources (No 7, 23, and 79) flux variability could be verified
by the HRI observations (see Sec.\ \ref{variability}). 

\subsubsection{X-ray binaries}

10 known HMXBs in the SMC were detected
both by the PSPC and by the HRI. Two additional HMXBs, AX J0051-722
and SMC X-3, were observed by the HRI only. 

Sources No 51 and 73 are HMXB candidates which were observed by
other X-ray satellites and are well known from literature.
Sources No 39, 81, and 88 were suggested as X-ray binaries by KPFH99.
KPFH99 additionally classified the sources No 12 and 83 either as
AGN or as XB candidates. For both there is a likely optical
counterpart within the HRI error circle with {\it B} = 16.8 and 17.3,
respectively. The hardness ratios are for No
12 HR1 = 1.00$\pm$0.61, HR2 = 0.43$\pm$0.16 and for No 83 HR1 =
1.00$\pm$1.30, HR2 = 1.00$\pm$1.93. 
10 HRI sources associated with emission line stars (Meyssonnier \&
Azzopardi 1993) in the SMC
were classified as new HMXB candidates forming Be/X-ray binaries by
Haberl \& Sasaki (2000, hereafter HS00). 

\subsubsection{AGN}

Source No 8 is a known AGN with the redshift z = 0.922 (Tinney \et\
1997). The hardness ratios are HR1 = 1.00$\pm$1.09, HR2 =
--0.06$\pm$0.12. The optical counterpart seen on DSS2 at the HRI
position, which is the same as reported by Tinney \et\ (1997), has {\it B}
= 18.6 (log(f$_{x}$/f$_{opt}$) = --0.49). 

Source No 59, showing both radio (SMC B0053-7227 (FHW98)) and X-ray
emission, was classified as an AGN candidate by KPFH99. A faint
optical source ({\it B} = 18.6) can be found at the HRI position
resulting in log(f$_{x}$/f$_{opt}$) = --0.30.
For the source No 112, also an AGN candidate (FHP00), no
optical counterpart was found. The PSPC pointing shows a hard source
(HR1 = 1.00$\pm$0.37, HR2 = 0.46$\pm$0.09).  
Seven HRI sources were classified as AGN candidates by KPFH99. For
them likely optical counterparts exist with brightness $17 < B
< 19$. The computed log(f$_{x}$/f$_{opt}$) are around --0.50.

\subsection{New classifications}\label{newclass}

In addition to the identification of HRI sources with known objects we 
classified new SMC X-ray
sources with the help of their properties obtained from HRI
observations. First we looked for extended sources which could suggest
new SNR candidates, but without any result. No new 
source with significant extent was detected. 

By comparing positions of X-ray sources with those of stars measured
in the optical and near-infrared one can conclude if the X-ray source
is in coincidence with a Galactic foreground star.
Therefore the HRI source catalogue was cross-correlated not only with the 
USNO-A2.0 catalogue, but also with the DENIS catalogue, looking 
for stars within the error circle of the X-ray detection.
Hardness ratios determined by PSPC are additional parameters for classifying 
hard X-ray sources not belonging to the Galaxy as candidates for XBs or AGN.

Table \ref{classified} summarizes the new classifications of this
work, and in Fig.\ \ref{fchart} DSS2 images around these sources
are presented with X-ray position and positional error.

\subsubsection{Sources classified as Galactic foreground
stars}\label{stellar} 

\begin{figure}[t]
\centerline{\psfig{figure=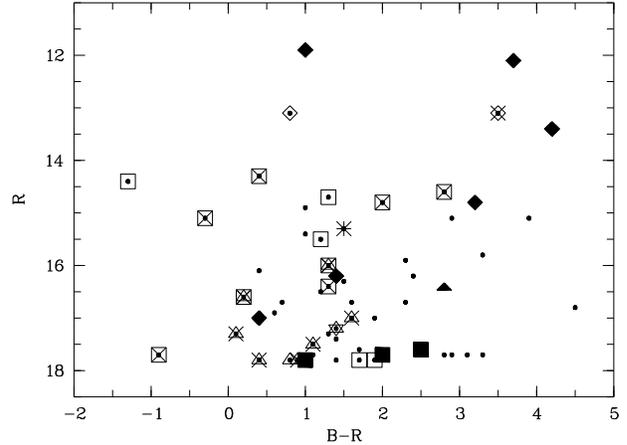,angle=270,width=9cm}}
\caption[]{\label{usno} Color-magnitude diagram for HRI sources 
correlating with entries in the USNO-A2.0 catalogue. 
Small dots are used for all sources. Extra asterisks indicate SSSs, 
stars SNRs, open squares XBs, open lozenges stars, and open triangles AGN.
Already known candidates are marked with crossed symbols, so an overlay of 
open triangle, open square, and cross marks hard sources as candidates
either for AGN or XB. Filled symbols are new classifications.}
\end{figure}

\begin{figure}[t]
\centerline{\psfig{figure=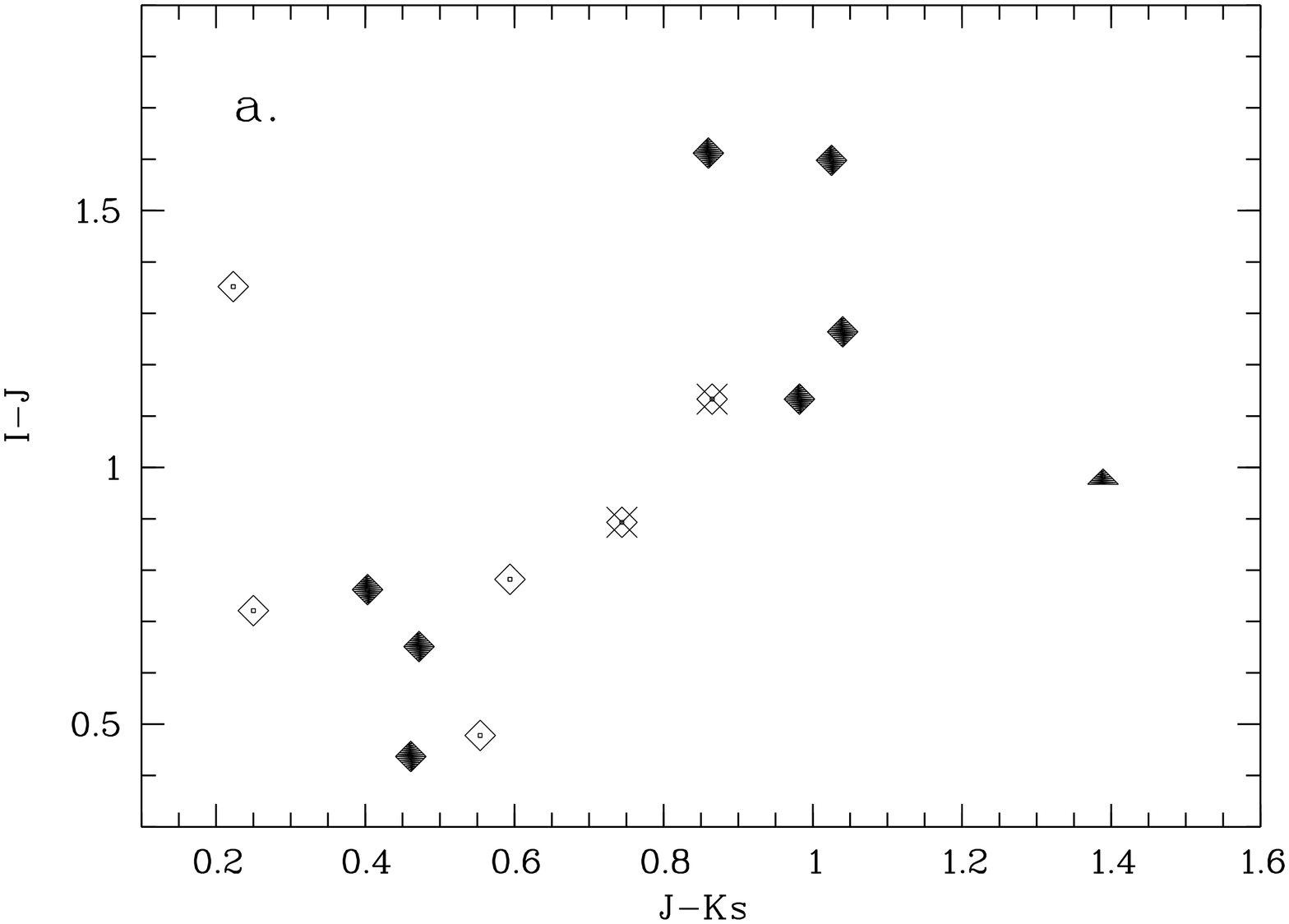,angle=270,width=9cm}}
\centerline{\psfig{figure=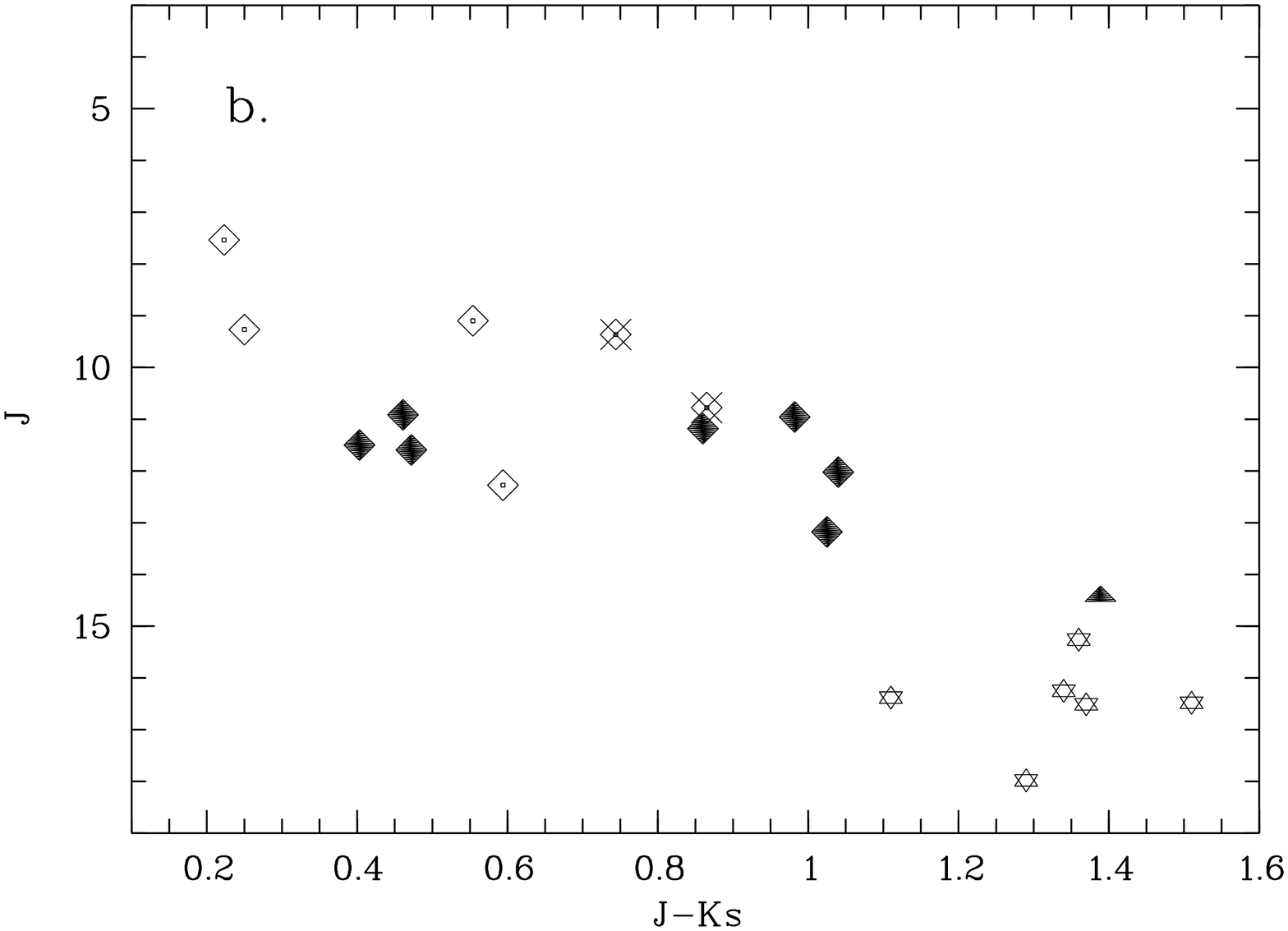,angle=270,width=9cm}}
\caption[]{\label{denis} Color-color and color-magnitude diagram 
for HRI sources correlating with objects in the DENIS catalogue. 
Open lozenges for foreground stars, crossed for candidates, and filled for
new classifications. The filled triangle mark the source classified as AGN. 
Stars signify AGN observed by Maiolino \et\ (2000).}
\end{figure}

In comparison to SMC members or background objects, foreground stars are 
brighter in {\it R} as shown in the color-magnitude diagram 
(Fig.\ \ref{usno}). For the near-infrared CLH00 estimated the number of
foreground stars and Magellanic Cloud members in the DENIS catalogue 
by studying color-color and color-magnitude diagrams for the DENIS data.
They showed that sources with $J - K_{s}$ lower than $\sim 1.0$ 
are most likely foreground stars with increasing probability for
smaller $J -K_{s}$ and smaller $I - J$ that such a classification is
correct. In our catalogue there are in total 14 sources correlating with
entries in the DENIS catalogue. Apart from one, $J - K_{s} < 1.1$ was 
measured for all of them.
Since sources observable both in near-infrared and
in X-rays are mainly galactic stars or AGN in the background, 
seven out of these 13 sources with $J - K_{s} < 1.1$ (No 16, 26,
35, 68, 80, 89, 121) were classified as foreground
stars. In Fig.\ \ref{denis}a they show a good correlation 
between $J - K_{s}$ and $I - J$ which is also shown by CLH00. 
For most of them we got $R <$ 14.0 from 
the USNO-A2.0 catalogue and likely optical counterparts can be seen in the
DSS2 images within the error circles (Fig.\ \ref{fchart}).
In addition the source No 25 was classified as a foreground star since an 
optical counterpart with {\it R} = 10.8 was found.

\subsubsection{Candidate for AGN}\label{newAGN}

In the color-color diagram in Fig.\ \ref{denis} there is an outlying source 
(No 64). Looking at the color-magnitude diagram (Fig.\ \ref{denis}b) it is 
the one with the highest $J - K_{s}$ and highest {\it J}. For comparison 
magnitudes and colors for a sample of AGN determined by Maiolino \et\ (2000)
are also plotted. The near-infrared and optical brightness of the HRI 
source No 68 correspond well with those typical for AGN ({\it B} = 19.2,
{\it R} = 16.4). So it was classified as an AGN.
 
\subsubsection{Candidates for AGN or X-ray binaries}\label{AGNorXB}

For the source No 71 no optical counterpart can be found on the DSS2
frame. In the USNO-A2.0 catalogue there is a faint object ({\it B} = 20.1,
{\it R} = 17.6) within the error circle, and with
log(f$_{x}$/f$_{opt}$) = 0.40 the HRI source is bright in X-rays. The
PSPC spectrum of this source is hard (HR1 = 1.00$\pm$0.13, HR2 =
0.38$\pm$0.09) favoring its classification as XB or AGN. 

Another five HRI sources (No 5, 53, 72, 78, 85) are associated with hard PSPC 
sources with HR1 = 1.00 and HR2 $\geq$ 0.25. For all of them large
errors were determined for HR1, but not for HR2 (HR2 -- error$_{\rm
HR2} >$ 0.00), resulting from the fact that the source is so hard that
the photon statistics were low in the softer  
bands (for definition of the hardness ratios see HFPK00). In the USNO-A2.0 
catalogue only faint or no optical source is found at the HRI position within 
the error circle. This can be also verified on the DSS2 images (see Fig.\ 
\ref{fchart}). Finally these HRI sources were classified as XB or AGN.

\section{Conclusion}

\begin{table}[t]
\caption[]{\label{numberlist} Number of classified X-ray sources in
the LMC and SMC observed by the HRI} 
\begin{tabular}{|l|rr|rr|}
\hline
\multicolumn{1}{|c|}{1} & 2 & 3 & 4 & 5 \\
\hline\hline
\multicolumn{1}{|c|}{Source} & \multicolumn{2}{c|}{LMC} &
\multicolumn{2}{c|}{SMC} \\
\multicolumn{1}{|c|}{class} & \multicolumn{1}{c}{id.+cand.} &
\multicolumn{1}{c|}{new} & \multicolumn{1}{c}{id.+cand.} &
\multicolumn{1}{c|}{new} \\ 
\hline\hline
SNR & 24 + 4 & 5 & 6 + 1 & \\
SSS & 5 & & 4 &  \\
XB & 9 + 1 & 2 & 12 + 15 & \\
AGN & 9 + 1 & & 1 + 9 & 1 \\
hard \scriptsize{(XB or AGN)} & & 3 & & 6 \\ 
foreground star & 39 + 2 & & 4 + 2 & 8 \\
\hline
\end{tabular}

\vspace{1mm}
Notes to columns No 2 and 4:

X-ray sources identified with known sources in catalogues and
literature + candidates reported in literature.

Notes to columns No 3 and 5:

New classifications of SHP00 and this work.
\end{table}

Analyzing the whole set of 71 ROSAT HRI pointings of the SMC region we
obtained a catalogue containing 121 discrete X-ray sources. Sources
were searched both in single pointings and merged data applying a
maximum likelihood algorithm. The high spatial resolution of the HRI
allowed us to identify X-ray sources with optically observed objects
in the TYCHO catalogue or the SIMBAD data base and to reduce the
systematic boresight errors. In total coordinates of 99 HRI sources
were corrected and the positional errors reduced to 5\farcs6 on
average.  

By searching through literature 56 HRI sources were identified with
already classified objects belonging to SMC SSSs, XBs,
SNRs, background AGN, or foreground stars.
Figure \ref{known} shows HRI sources which are identified with 
objects of different source classes known from various 
catalogues and literature, plotted over an 0.1 -- 2.4 keV PSPC
image. 
The HRI observations mostly covered the main body of the
SMC (see Fig.\ \ref{exposure}), which is the reason that most of the 
identified and new sources are located in this region (see Fig.\ 
\ref{unidentified}). 
In this work 15 sources are newly classified as foreground star, 
XB, or AGN.  
There are still 50 unclassified HRI detections well distributed 
over the field observed by the HRI. The source distribution 
shows that there is a deficiency of X-ray sources in the central part of 
the SMC around RA = 00$^h$ 55$^m$, Dec = --72$\degr\ $ 40$\arcmin\ $
(J2000.0).

In Table \ref{numberlist} the numbers of detected sources are compared
between the LMC and the SMC. 
In the LMC the number of SNRs is larger
than in the SMC by a factor of four roughly in accordance with the ratio 
of the areas covered by the observations.
The number of known SSSs and X-ray binaries detected in ROSAT observations 
is comparable in the two clouds, most of the XBs are binary 
systems with Be-star companions. Nevertheless the number of XB candidates 
in the SMC is relatively higher than in the LMC maybe partly due to the fact 
that a thorough search for Be/X-ray binary candidates has been done for the
SMC by comparing the X-ray catalogues with the catalogues of emission line
objects (HS00). 

Based on all available ROSAT HRI pointing data complete HRI source
catalogues both for the LMC and for the SMC are now available. They
summarize the detections of X-ray sources in the MCs and provide
accurate positions with errors smaller than 16\arcsec\ for all
sources, for the majority of sources about 8\arcsec\,. Verification of
the candidates for the different source classes by a final optical
identification is still needed. Additional observations
in the future will achieve more precise characterization of the sources
giving a better overall picture of the MCs.   

\acknowledgements
The ROSAT project is supported by the German
Bundesministerium f\"ur Bildung und Forschung (BMBF) and 
the Max-Planck-Gesellschaft. This research has been carried out by
making extensive use of the SIMBAD data base operated at CDS,
Strasbourg, France, and data obtained through the High Energy Astrophysics 
Science Archive Research Center Online Service, provided by the NASA/Goddard 
Space Flight Center. 

\onecolumn

\begin{figure}
\caption[]{\label{fchart} Positions of newly classified HRI sources on
DSS2 images. The comments in brackets $<$ $>$ give the classifications. 
The X-ray source position is in the center of the image, circles mark
the 90 \% confidence error.}
\caption[]{\label{known} Distribution of identified HRI sources in the
SMC region plotted on a grey scale image (0.1 -- 2.4 keV) obtained
from the PSPC data. 
Squares are used for SNRs, double squares for SSSs, crossed squares
for XBs, crossed circles for AGN, circles for foreground
stars. Candidates from literature are included for each source class.} 
\caption[]{\label{unidentified} The distribution of unidentified HRI
sources and new source classifications is shown. Unidentified HRI
sources are plotted as small squares, circles are classifications as
foreground star, crossed circle as AGN, and double circles as AGN or XB 
candidates.} 
\end{figure}

\clearpage

\begin{landscape}

\begin{table}
\scriptsize

\caption[]{Identified HRI sources in the SMC}
\begin{tabular}{rrrrrccrrccl}
\hline\noalign{\smallskip}
1~ & \multicolumn{1}{c}{2} & \multicolumn{1}{c}{3} & 4~ & \multicolumn{1}{c}{5} & 6 & 7 & \multicolumn{1}{c}{8} & \multicolumn{1}{c}{9} & 10 & 11 & ~~~12 \\
\hline\noalign{\smallskip}  
No & \multicolumn{1}{c}{RA} & \multicolumn{1}{c}{Dec} & \perr & \exil
& Count rate & \ext & \extl & \multicolumn{1}{c}{No} & HR1 & HR2 & Remarks \\
   & \multicolumn{2}{c}{(J2000.0)} & [\arcsec] & & [\ct] & [\arcsec] &
& \multicolumn{1}{c}{PSPC} & & & \\ 
\noalign{\smallskip}\hline\noalign{\smallskip}
 10 & 00 40 00.6 & -73 45 42 &  8.8 &    33.7 & 3.18e-03$\pm$5.26e-04 &  0.0$\pm$ 0.0 &   0.0 &  562 & -0.05$\pm$0.07 &  0.05$\pm$0.10 & foreground star HD 3880                                                          \\ 
 18 & 00 46 41.8 & -73 01 15 &  5.6 &    23.9 & 1.77e-03$\pm$4.42e-04 &  0.0$\pm$ 0.0 &   0.0 &  383 & -0.26$\pm$0.10 & -0.39$\pm$0.15 & foreground star? [HFPK00]                                                         \\ 
 86 & 01 00 13.3 & -73 07 24 &  7.2 &   234.9 & 6.94e-03$\pm$6.05e-04 &  0.0$\pm$ 0.0 &   0.0 &  408 & -0.27$\pm$0.10 &  0.23$\pm$0.15 & foreground star? [KPFH99]                                                         \\ 
 92 & 01 00 56.2 & -72 33 54 &  4.7 &    14.7 & 7.57e-04$\pm$2.00e-04 &  2.4$\pm$ 1.9 &   1.9 &  273 &  1.00$\pm$1.61 &  1.00$\pm$0.47 & foreground star HD 6171                                                          \\ 
119 & 01 17 28.9 & -73 11 44 &  4.2 &    12.0 & 2.24e-03$\pm$7.85e-04 &  2.3$\pm$ 2.4 &   0.9 &  431 &  1.00$\pm$0.54 &  0.30$\pm$0.09 & foreground star G [SCC99]                                                        \\ 
120 & 01 18 37.9 & -73 25 27 &  0.9 &  1546.7 & 2.74e-02$\pm$1.21e-03 &  2.9$\pm$ 1.1 &  40.5 &  478 &  0.12$\pm$0.02 &  0.07$\pm$0.02 & foreground star HD 8191 [CSM97]                                                  \\ 
\noalign{\smallskip}
 22 & 00 47 40.5 & -73 09 28 &  6.3 &    11.7 & 1.74e-03$\pm$5.63e-04 &  3.6$\pm$ 2.8 &   1.4 &  419 &  1.00$\pm$0.09 &  0.27$\pm$0.04 & SNR? [KPFH99]                                                                     \\ 
 38 & 00 51 02.4 & -73 21 31 &  2.7 &    50.2 & 5.49e-03$\pm$6.58e-04 &  6.2$\pm$ 2.3 &  56.6 &  461 &  0.79$\pm$0.01 & -0.26$\pm$0.02 & SNR 0049-73.6                                                                    \\ 
 77 & 00 58 16.0 & -72 18 04 &  4.5 &    69.4 & 3.00e-03$\pm$3.56e-04 &  3.7$\pm$ 3.0 &   1.3 &  194 &  1.00$\pm$0.14 &  0.15$\pm$0.07 & SNR 0056-72.5                                                                    \\ 
 82 & 00 59 26.2 & -72 10 02 &  4.0 &    30.5 & 3.10e-03$\pm$3.37e-04 &  5.8$\pm$ 2.0 &  56.5 &  148 &  0.90$\pm$0.02 & -0.12$\pm$0.03 & SNR 0057-72.2                                                                    \\ 
107 & 01 04 00.3 & -72 02 03 &  3.3 &  9732.8 & 2.47e-01$\pm$2.75e-03 &  5.8$\pm$ 0.8 &8561.7 &  107 &  0.92$\pm$0.00 & -0.20$\pm$0.01 & SNR 0102-72.3                                                                    \\ 
109 & 01 05 02.6 & -72 22 57 &  3.6 &    92.7 & 6.71e-03$\pm$5.16e-04 &  6.3$\pm$ 1.8 & 161.0 &  217 &  0.66$\pm$0.01 & -0.32$\pm$0.02 & SNR 0103-72.6                                                                    \\ 
113 & 01 06 14.9 & -72 05 27 &  7.4 &    38.1 & 8.49e-03$\pm$9.16e-04 & 15.5$\pm$ 5.3 &  35.5 &  125 &  1.00$\pm$0.06 &  0.19$\pm$0.04 & SNR 0104-72.3                                                                    \\ 
\noalign{\smallskip}
  7 & 00 37 19.8 & -72 14 13 &  1.1 & 10945.0 & 8.14e-02$\pm$1.54e-03 &  3.2$\pm$ 0.6 &1066.7 &  176 & -0.97$\pm$0.00 & -1.00$\pm$0.06 & SSS 1E0035.4-7230                                                                \\ 
 23 & 00 48 19.6 & -73 31 52 &  2.3 &  5165.4 & 6.56e-02$\pm$1.99e-03 &  1.2$\pm$ 0.5 &  12.8 &  512 & -1.00$\pm$0.24 &     & SSS RXJ0048.4-7332                                                               \\ 
 79 & 00 58 37.3 & -71 35 49 &  1.5 &   700.3 & 5.17e-02$\pm$3.77e-03 &  2.3$\pm$ 1.0 &  29.0 &   47 & -0.99$\pm$0.00 & -1.00$\pm$0.26 & SSS 1E0056.8-7146                                                                \\ 
106 & 01 03 53.9 & -72 54 41 &  7.3 &   197.6 & 6.00e-03$\pm$5.65e-04 &  0.0$\pm$ 0.0 &   0.0 &  361 & -1.00$\pm$0.10 &     & SSS RXJ0103.8-7254 [KP96],[KPFH99]                                                \\ 
\noalign{\smallskip}
 12 & 00 42 41.8 & -73 40 41 &  6.7 &    25.7 & 1.00e-03$\pm$2.30e-04
&  1.8$\pm$ 1.7 &   1.0 &  546 &  1.00$\pm$0.61 &  0.43$\pm$0.16 &
AGN? or XB? [KPFH99] \\ 
 28 & 00 49 30.8 & -73 31 09 &  3.0 &    69.4 & 2.01e-03$\pm$2.94e-04 &  2.7$\pm$ 1.7 &   7.0 &  511 &  1.00$\pm$0.45 &  0.32$\pm$0.17 & HMXB? Be/X [HS00]                                                           \\ 
 34 & 00 50 42.9 & -73 16 08 &  1.8 &   126.2 & 4.23e-03$\pm$5.53e-04 &  1.6$\pm$ 1.4 &   1.0 &  444 &  1.00$\pm$0.24 &  0.58$\pm$0.10 & HMXB Be star in SMC                                                              \\ 
 36 & 00 50 57.2 & -73 10 06 &  3.3 &    53.4 & 2.24e-03$\pm$3.52e-04 &  0.8$\pm$ 2.8 &  0.0 &  421 &  1.00$\pm$0.33 &  0.52$\pm$0.12 & HMXB? Be/X [HS00]                                                           \\ 
 37 & 00 50 59.3 & -72 13 26 &  2.6 &   115.1 & 2.11e-02$\pm$3.71e-03 &  0.4$\pm$ 1.1 &   0.0 &    &     &     & HMXB Be/X AXJ0051-722 [KP96]                                                     \\ 
 39 & 00 51 20.5 & -72 16 41 &  4.6 &    10.1 & 3.53e-03$\pm$1.59e-03 &  0.0$\pm$ 0.0 &   0.0 &  188 &  1.00$\pm$0.80 &  1.00$\pm$0.56 & XB? [KPFH99]                                                                      \\ 
 41 & 00 51 52.0 & -73 10 32 &  1.5 &   193.6 & 3.55e-03$\pm$3.83e-04 &  0.5$\pm$ 1.1 &   0.0 &  424 &  1.00$\pm$0.08 &  0.36$\pm$0.06 & HMXB Be/X RXJ0051.9-7311 [CSM97],[SCC99],[HS00]                                    \\ 
 43 & 00 52 07.7 & -72 25 50 & 15.9 &    10.3 & 1.19e-02$\pm$3.69e-03 &  6.3$\pm$ 9.8 &   0.1 &    &     &     & HMXB SMC X-3                                                                     \\ 
 44 & 00 52 13.9 & -73 19 18 &  1.1 & 23375.0 & 6.79e-01$\pm$5.10e-03 &  1.1$\pm$ 0.6 &   4.5 &  453 &  1.00$\pm$0.07 &  0.61$\pm$0.04 & HMXB Be/X RXJ0052.1-7319, pulsar [LPM99],[ISC99]                                 \\ 
 46 & 00 52 55.2 & -71 58 06 &  3.0 &    83.6 & 2.37e-03$\pm$3.27e-04 &  2.9$\pm$ 1.9 &   3.9 &   94 &  1.00$\pm$0.43 &  0.44$\pm$0.13 & HMXB Be/X RXJ0052.9-7158 [CSM97],[SCC99],[HS00]                                   \\ 
 48 & 00 53 24.1 & -72 27 20 &  3.1 &    39.3 & 1.23e-03$\pm$2.40e-04 &  1.3$\pm$ 1.6 &   0.2 &  246 &  1.00$\pm$0.59 &  1.00$\pm$1.29 & HMXB? Be/X [HS00]                                                           \\ 
 51 & 00 53 56.0 & -72 26 52 &  2.4 &   395.2 & 7.76e-03$\pm$5.59e-04 &  3.1$\pm$ 1.2 &  56.3 &  242 &  1.00$\pm$0.18 &  0.46$\pm$0.05 & HMXB? XTEJ0053-724, pulsar [CML98]                                               \\ 
 57 & 00 54 56.0 & -72 45 06 &  9.7 &    34.7 & 3.02e-03$\pm$4.84e-04 &  0.0$\pm$ 5.2 &   0.0 &  324 &  1.00$\pm$0.71 &  0.58$\pm$0.14 & HMXB? Be/X [HS00]                                                           \\ 
 58 & 00 54 56.5 & -72 26 52 &  2.6 &   127.8 & 2.82e-03$\pm$3.49e-04 &  1.5$\pm$ 1.5 &   0.5 &  241 &  1.00$\pm$0.04 &  0.56$\pm$0.03 & HMXB Be/X XTEJ0055-724 [SCB99]                            \\ 
 73 & 00 57 48.5 & -72 02 34 &  6.4 &    22.9 & 3.08e-03$\pm$6.79e-04 &  0.0$\pm$ 0.0 &   0.0 &  114 &  1.00$\pm$0.52 &  1.00$\pm$0.35 & HMXB? AXJ0058-72.0, pulsar [YK98a]                                               \\ 
 74 & 00 57 49.3 & -72 07 55 &  3.5 &   236.9 & 4.51e-03$\pm$3.66e-04 &  1.2$\pm$ 1.5 &   0.2 &  136 &  1.00$\pm$0.22 &  0.58$\pm$0.11 & HMXB? Be/X [HS00]                                                           \\ 
 76 & 00 58 12.6 & -72 30 50 &  4.5 &   103.7 & 4.28e-03$\pm$4.77e-04 &  0.0$\pm$ 0.0 &   0.0 &  258 &  1.00$\pm$1.06 &  1.00$\pm$0.89 & HMXB Be/X [SCC99]                                                                  \\ 
 81 & 00 59 21.5 & -72 23 20 &  5.0 &    11.4 & 1.43e-03$\pm$5.84e-04 &  1.1$\pm$ 2.0 &   0.0 &  218 &  1.00$\pm$0.41 &  0.52$\pm$0.06 & XB? [KPFH99]                                                                      \\ 
 83 & 01 00 04.6 & -71 57 24 &  8.9 &    12.6 & 1.50e-03$\pm$3.53e-04
&  0.0$\pm$ 0.0 &   0.0 &   91 &  1.00$\pm$1.30 &  1.00$\pm$1.93 &
AGN? or XB? [KPFH99]\\ 
 88 & 01 00 15.0 & -72 04 41 &  4.1 &    48.4 & 1.66e-03$\pm$2.47e-04 &  2.7$\pm$ 2.0 &   1.9 &  123 &  1.00$\pm$0.38 &  1.00$\pm$0.57 & XB? [KPFH99]                                                                      \\ 
 93 & 01 01 02.1 & -72 06 56 &  3.7 &   181.6 & 1.15e-02$\pm$1.31e-03 &  2.3$\pm$ 1.8 &   1.7 &  132 &  1.00$\pm$0.15 &  0.47$\pm$0.04 & HMXB Be/X RXJ0101.0-7206 [KP96],[SCB99]                                          \\ 
 95 & 01 01 19.5 & -72 11 18 &  3.6 &   235.1 & 5.61e-03$\pm$4.24e-04 &  3.4$\pm$ 1.9 &   5.9 &  159 &  1.00$\pm$0.98 &  1.00$\pm$0.47 & HMXB? Be/X [HS00]                                                           \\ 
 96 & 01 01 36.4 & -72 04 18 &  4.3 &    96.4 & 3.91e-03$\pm$4.05e-04 &  0.0$\pm$ 0.0 &   0.0 &  121 &  1.00$\pm$0.29 &  0.18$\pm$0.09 & HMXB? Be/X [HS00]                                                           \\ 
 97 & 01 01 53.3 & -72 23 35 & 10.3 &    15.2 & 1.95e-03$\pm$4.45e-04 &  0.0$\pm$ 0.0 &   0.0 &  220 &  1.00$\pm$0.28 &  0.68$\pm$0.08 & HMXB? Be/X [HS00]                                                           \\ 
101 & 01 03 13.3 & -72 09 17 &  3.8 &    48.4 & 2.30e-03$\pm$4.40e-04 &  1.8$\pm$ 1.7 &   1.1 &  143 &  1.00$\pm$0.10 &  0.24$\pm$0.06 & HMXB SAXJ0103.2-7209, pulsar [ISC98],[HS94]                           \\ 
105 & 01 03 37.2 & -72 01 35 &  3.6 &    53.9 & 1.72e-03$\pm$3.07e-04 &  0.6$\pm$ 1.3 &   0.0 &  106 &  1.00$\pm$7.13 &  0.41$\pm$0.07 & HMXB? Be/X [HS00]                                                           \\ 
108 & 01 04 35.9 & -72 21 48 &  4.0 &    21.9 & 9.63e-04$\pm$2.63e-04
&  0.0$\pm$ 0.0 &   0.0 &    &     &     & HMXB? Be/X [HS00]                                                           \\ 
110 & 01 05 09.3 & -72 11 50 &  7.8 &    13.2 & 1.58e-03$\pm$3.99e-04 &  0.0$\pm$ 0.0 &   0.0 &  163 &  1.00$\pm$0.20 &  0.40$\pm$0.09 & HMXB AXJ0105-722, pulsar [YK98b],[FHP00]                                         \\ 
118 & 01 17 05.1 & -73 26 37 &  0.7 & 26285.8 & 3.12e+00$\pm$5.42e-02 &  1.9$\pm$ 0.4 & 289.2 &  482 &  0.97$\pm$0.00 &  0.39$\pm$0.00 & HMXB SMC X-1                                                                     \\ 
\noalign{\smallskip}
\hline
\end{tabular}

\vspace{2mm}
Notes to column No 9 (Table \ref{identified}), No 10 (Tables
\ref{classified} and \ref{wholecat}): 
Catalogue number from [HFPK00].

\vspace{1mm}
Notes to column No 12 (Table \ref{identified}), No 13
(Tables \ref{classified} and \ref{wholecat}):
Candidates from literature are marked with ? behind the source class. 
New classification of this work are put in $<$ $>$.

Abbreviations for references in square brackets are given in the
literature list.
\end{table}

\clearpage

\addtocounter{table}{-1}
\begin{table}
\scriptsize
\caption[]{Continued}
\begin{tabular}{rrrrrccrrccl}
\hline\noalign{\smallskip}
1~ & \multicolumn{1}{c}{2} & \multicolumn{1}{c}{3} & 4~ & \multicolumn{1}{c}{5} & 6 & 7 & \multicolumn{1}{c}{8} & \multicolumn{1}{c}{9} & 10 & 11 & ~~~12 \\
\hline\noalign{\smallskip}  
No & \multicolumn{1}{c}{RA} & \multicolumn{1}{c}{Dec} & \perr & \exil & Count rate & \ext & \extl & \multicolumn{1}{c}{No} & HR1 & HR2 & Remarks \\
   & \multicolumn{2}{c}{(J2000.0)} & [\arcsec] & & [\ct] & [\arcsec] &
& \multicolumn{1}{c}{PSPC} & & & \\ 
\noalign{\smallskip}\hline\noalign{\smallskip}
  3 & 00 35 52.7 & -72 09 19 &  6.0 &    12.3 & 8.57e-04$\pm$2.27e-04 &  2.9$\pm$ 3.2 &   0.3 &  144 &  1.00$\pm$2.57 &  0.33$\pm$0.23 & AGN? [KPFH99]                                                                     \\ 
  4 & 00 36 01.8 & -72 21 12 &  2.2 &   152.8 & 4.12e-03$\pm$3.99e-04 &  2.5$\pm$ 2.1 &   1.2 &  211 &  0.35$\pm$0.13 & -0.04$\pm$0.13 & AGN? [KPFH99]                                                                     \\ 
  6 & 00 36 11.4 & -72 17 43 &  4.2 &    16.6 & 8.56e-04$\pm$2.06e-04 &  2.9$\pm$ 2.7 &   0.8 &  191 &  1.00$\pm$0.76 &  0.10$\pm$0.18 & AGN? [KPFH99]                                                                     \\ 
  8 & 00 37 20.5 & -72 18 05 &  1.9 &    74.6 & 1.66e-03$\pm$2.41e-04 &  1.7$\pm$ 1.4 &   1.7 &  193 &  1.00$\pm$1.09 & -0.06$\pm$0.12 & AGN z=0.922 [TDZ97]                                                              \\ 
 14 & 00 43 56.8 & -73 42 20 &  7.2 &    11.4 & 6.47e-04$\pm$1.98e-04 &  1.9$\pm$ 1.9 &   0.8 &  552 &  1.00$\pm$0.99 &  1.00$\pm$1.99 & AGN? [KPFH99]                                                                     \\ 
 59 & 00 55 28.7 & -72 10 55 &  3.8 &    25.9 & 2.58e-03$\pm$6.66e-04 &  1.6$\pm$ 1.9 &   0.2 &  157 &  1.00$\pm$0.04 &  0.36$\pm$0.04 & AGN? Radio SMC B0053-7227 [FHW98],[KPFH99]                                             \\ 
 90 & 01 00 42.2 & -72 11 33 &  3.3 &  1610.5 & 1.59e-02$\pm$6.34e-04 &  2.1$\pm$ 0.9 &  22.5 &  162 &  1.00$\pm$0.07 &  0.33$\pm$0.04 & AGN? [KPFH99]                                                                     \\ 
 98 & 01 02 41.8 & -72 32 40 &  4.9 &    44.1 & 2.19e-03$\pm$3.48e-04 &  0.0$\pm$ 0.0 &   0.0 &  267 &  1.00$\pm$0.33 &  0.49$\pm$0.08 & AGN? [KPFH99]                                                                     \\ 
104 & 01 03 28.8 & -72 47 27 &  9.7 &    40.2 & 3.88e-03$\pm$5.84e-04 &  0.0$\pm$ 0.0 &   0.0 &  334 &  1.00$\pm$0.33 & -0.08$\pm$0.07 & AGN? [KPFH99]                                                                     \\ 
112 & 01 05 33.7 & -72 13 29 &  4.7 &    40.6 & 1.76e-03$\pm$2.99e-04 &  0.0$\pm$ 0.0 &   0.0 &  172 &  1.00$\pm$0.37 &  0.46$\pm$0.09 & AGN? [FHP00]                                                                     \\ 
\noalign{\smallskip}
\hline
\end{tabular}
\label{identified}

\caption[]{
Classified HRI sources
}
\begin{tabular}{rrrrrccrrrccl}
\hline\noalign{\smallskip}
1~ & \multicolumn{1}{c}{2} & \multicolumn{1}{c}{3} & 4~ &
\multicolumn{1}{c}{5} & 6 & 7 & \multicolumn{1}{c}{8} &
\multicolumn{1}{c}{9} & \multicolumn{1}{c}{10} & 11 & 12 & ~~~13 \\
\hline\noalign{\smallskip}  
No & \multicolumn{1}{c}{RA} & \multicolumn{1}{c}{Dec} & \perr & \exil
& Count rate & \ext & \extl & log(f$_{x}$/f$_{opt}$) & \multicolumn{1}{c}{No} & HR1 & HR2 & Remarks \\
   & \multicolumn{2}{c}{(J2000.0)} & [\arcsec] & & [\ct] & [\arcsec] &
& & \multicolumn{1}{c}{PSPC} & & & \\ 
\noalign{\smallskip}\hline\noalign{\smallskip}
  5 & 00 36 04.7 & -72 18 44 &  3.8 &    13.1 & 6.09e-04$\pm$1.72e-04 &  0.0$\pm$ 0.0 &   0.0 &-0.85 &  198 &  1.00$\pm$3.45 &  1.00$\pm$0.49 & $<$XB$>$ or $<$AGN$>$                                                           \\
 16 & 00 45 10.9 & -73 04 17 &  7.5 &    10.8 & 1.69e-03$\pm$5.81e-04 &  0.0$\pm$ 0.0 &   0.0 &-1.61 &  397 &  1.00$\pm$0.88 & -0.18$\pm$0.12 & $<$foreground star$>$ GSC 9141.7926                                             \\  
 25 & 00 48 55.6 & -73 33 37 &  3.3 &    28.7 & 1.18e-03$\pm$2.92e-04 &  0.0$\pm$ 0.0 &   0.0 &-3.68 &      &      &      & $<$foreground star$>$ GSC 9141.7623                                             \\  
 26 & 00 48 59.0 & -72 58 13 &  5.3 &    18.9 & 1.03e-03$\pm$2.50e-04 &  0.0$\pm$ 0.0 &   0.0 & &  &  &  & $<$foreground star$>$                                                           \\  
 35 & 00 50 49.3 & -72 41 53 &  4.3 &   447.9 & 1.33e-02$\pm$8.11e-04 &  0.0$\pm$ 0.0 &   0.0 & 0.01 &      &      &      & $<$foreground star$>$ SkKM 62                                                   \\  
 53 & 00 54 32.1 & -72 18 13 &  3.7 &    27.4 & 1.23e-03$\pm$2.56e-04 &  2.4$\pm$ 2.2 &   0.9 & &  196 &  1.00$\pm$1.37 &  1.00$\pm$0.59 & $<$XB$>$ or $<$AGN$>$                                                           \\  
 64 & 00 55 49.3 & -72 28 37 &  6.5 &    10.1 & 1.03e-03$\pm$3.35e-04 &  0.0$\pm$ 0.0 &   0.0 &-0.46 &      &      &      & $<$AGN$>$                                                                       \\  
 68 & 00 56 50.3 & -73 10 38 &  9.9 &    10.0 & 9.85e-04$\pm$2.90e-04 &  0.0$\pm$ 0.0 &   0.0 &-3.00 &  426 &  1.00$\pm$0.38 &  0.34$\pm$0.15 & $<$foreground star$>$ GSC 9141.7719                                             \\  
 71 & 00 57 31.7 & -72 13 01 &  3.9 &   111.0 & 3.26e-03$\pm$3.40e-04 &  0.0$\pm$ 0.0 &   0.0 & 0.40 &  170 &  1.00$\pm$0.13 &  0.38$\pm$0.09 & $<$XB$>$ or $<$AGN$>$                                                           \\  
 72 & 00 57 35.5 & -73 12 56 &  9.2 &    14.3 & 1.24e-03$\pm$3.04e-04 &  5.0$\pm$ 3.8 &   1.6 & &  435 &  1.00$\pm$0.84 &  0.25$\pm$0.16 & $<$XB$>$ or $<$AGN$>$                                                           \\  
 78 & 00 58 20.7 & -72 16 18 &  6.3 &    10.9 & 6.78e-04$\pm$1.96e-04 &  2.3$\pm$ 3.0 &   0.3 & &  185 &  1.00$\pm$0.81 &  0.58$\pm$0.09 & $<$XB$>$ or $<$AGN$>$                                                           \\  
 80 & 00 59 13.6 & -72 22 25 &  8.9 &    12.3 & 1.42e-03$\pm$3.37e-04 &  0.0$\pm$ 0.0 &   0.0 &-0.80 &  &  &  & $<$foreground star$>$                                                           \\  
 85 & 01 00 12.0 & -72 20 17 &  5.3 &    10.3 & 1.35e-03$\pm$5.71e-04 &  0.0$\pm$ 0.0 &   0.0 &-0.14 &  204 &  1.00$\pm$0.99 &  1.00$\pm$0.58 & $<$XB$>$ or $<$AGN$>$                                                           \\  
 89 & 01 00 36.5 & -73 00 35 &  7.1 &   164.9 & 3.64e-03$\pm$4.18e-04 &  1.2$\pm$ 1.2 &   0.5 &-0.63 &  & &  & $<$foreground star$>$                                                           \\  
121 & 01 19 39.2 & -73 27 32 &  7.0 &    13.8 & 1.50e-03$\pm$3.90e-04 &  0.0$\pm$ 0.0 &   0.0 &-0.94 &  488 & -0.19$\pm$0.15 & -1.00$\pm$1.95 & $<$foreground star$>$    \\
\noalign{\smallskip}
\hline
\end{tabular}
\label{classified} 

\end{table}

\clearpage

\begin{table}
\scriptsize
\caption[]{\label{wholecat}HRI sources in the SMC region}
\begin{tabular}{rrrrrccrrrccl}
\hline\noalign{\smallskip}
1~ & \multicolumn{1}{c}{2} & \multicolumn{1}{c}{3} & 4~ &
\multicolumn{1}{c}{5} & 6 & 7 & \multicolumn{1}{c}{8} &
\multicolumn{1}{c}{9} & \multicolumn{1}{c}{10} & 11 & 12 & ~~~13 \\
\hline\noalign{\smallskip}  
No & \multicolumn{1}{c}{RA} & \multicolumn{1}{c}{Dec} & \perr & \exil
& Count rate & \ext & \extl & Count rate & \multicolumn{1}{c}{No} & HR1 & HR2 & Remarks \\
 & & & & & & & & \multicolumn{1}{c}{PSPC} & \multicolumn{1}{c}{PSPC} & & & \\ 
 & \multicolumn{2}{c}{(J2000.0)} & [\arcsec] & & [\ct] & [\arcsec] &
& \multicolumn{1}{c}{[\ct]} & & & & \\ 
\noalign{\smallskip}\hline\noalign{\smallskip}
  1 & 00 34 56.0 & -72 16 48 &  7.8 &    10.9 & 1.02e-03$\pm$2.77e-04 &  0.0$\pm$ 0.0 &   0.0 & 3.12E-03    & 187 &  1.00$\pm$1.34 &  1.00$\pm$0.89 &                                                                \\ 
  2 & 00 35 21.3 & -72 12 43 &  4.8 &    23.0 & 1.40e-03$\pm$2.77e-04 &  3.5$\pm$ 3.2 &   1.0 & 6.48E-03    & 167 &  1.00$\pm$0.46 &  0.30$\pm$0.12 & \\
  3 & 00 35 52.7 & -72 09 19 &  6.0 &    12.3 & 8.57e-04$\pm$2.27e-04 &  2.9$\pm$ 3.2 &   0.3 & 3.03E-03    & 144 &  1.00$\pm$2.57 &  0.33$\pm$0.23 & AGN? [KPFH99]                                                   \\ 
  4 & 00 36 01.8 & -72 21 12 &  2.2 &   152.8 & 4.12e-03$\pm$3.99e-04 &  2.5$\pm$ 2.1 &   1.2 & 7.34E-03    & 211 &  0.35$\pm$0.13 & -0.04$\pm$0.13 & AGN? [KPFH99]                                                   \\ 
  5 & 00 36 04.7 & -72 18 44 &  3.8 &    13.1 & 6.09e-04$\pm$1.72e-04 &  0.0$\pm$ 0.0 &   0.0 & 1.83E-03    & 198 &  1.00$\pm$3.45 &  1.00$\pm$0.49 & $<$XB$>$ or $<$AGN$>$                                                               \\ 
  6 & 00 36 11.4 & -72 17 43 &  4.2 &    16.6 & 8.56e-04$\pm$2.06e-04 &  2.9$\pm$ 2.7 &   0.8 & 3.68E-03    & 191 &  1.00$\pm$0.76 &  0.10$\pm$0.18 & AGN? [KPFH99]                                                   \\ 
  7 & 00 37 19.8 & -72 14 13 &  1.1 & 10945.0 & 8.14e-02$\pm$1.54e-03 &  3.2$\pm$ 0.6 &1066.7 & 4.40E-01    & 176 & -0.97$\pm$0.00 & -1.00$\pm$0.06 & SSS 1E0035.4-7230                                              \\ 
  8 & 00 37 20.5 & -72 18 05 &  1.9 &    74.6 & 1.66e-03$\pm$2.41e-04 &  1.7$\pm$ 1.4 &   1.7 & 7.49E-03    & 193 &  1.00$\pm$1.09 & -0.06$\pm$0.12 & AGN z=0.922 [TDZ97]                                            \\ 
  9 & 00 38 55.3 & -72 05 37 &  4.1 &    54.4 & 2.73e-03$\pm$3.76e-04 &  1.2$\pm$ 3.3 &  0.0 & 4.57E-03    & 127 &  1.00$\pm$0.56 &  0.32$\pm$0.15 & \\
 10 & 00 40 00.6 & -73 45 42 &  8.8 &    33.7 & 3.18e-03$\pm$5.26e-04 &  0.0$\pm$ 0.0 &   0.0 & 1.66E-02    & 562 & -0.05$\pm$0.07 &  0.05$\pm$0.10 & foreground star HD 3880                                        \\ 
 11 & 00 42 08.4 & -73 45 02 &  7.2 &    14.8 & 7.94e-04$\pm$2.21e-04 &  2.3$\pm$ 2.4 &   0.6 & $<$1.18E-03 &    &      &      &                                                                \\ 
 12 & 00 42 41.8 & -73 40 41 &  6.7 &    25.7 & 1.00e-03$\pm$2.30e-04 &  1.8$\pm$ 1.7 &   1.0 & 2.87E-03    & 546 &  1.00$\pm$0.61 &  0.43$\pm$0.16 & XB? or AGN? [KPFH99]                                            \\ 
 13 & 00 42 52.2 & -73 25 38 &  7.9 &    11.4 & 9.70e-04$\pm$3.55e-04 &  0.0$\pm$ 0.0 &   0.0 & $<$6.17E-03 &    &      &      &                                                                \\ 
 14 & 00 43 56.8 & -73 42 20 &  7.2 &    11.4 & 6.47e-04$\pm$1.98e-04 &  1.9$\pm$ 1.9 &   0.8 & 1.41E-03    & 552 &  1.00$\pm$0.99 &  1.00$\pm$1.99 & AGN? [KPFH99]                                                   \\ 
 15 & 00 44 56.9 & -73 22 38 &  5.7 &    10.2 & 2.31e-03$\pm$1.03e-03 &  0.0$\pm$ 0.0 &   0.0 &         &    &      &      &                                                                \\ 
 16 & 00 45 10.9 & -73 04 17 &  7.5 &    10.8 & 1.69e-03$\pm$5.81e-04 &  0.0$\pm$ 0.0 &   0.0 & 6.36E-03    & 397 &  1.00$\pm$0.88 & -0.18$\pm$0.12 & $<$foreground star$>$ GSC 9141.7926\\
 17 & 00 45 24.7 & -73 28 58 &  7.9 &    18.0 & 1.71e-03$\pm$4.89e-04 &  0.0$\pm$ 0.0 &   0.0 & 3.62E-03    & 494 &  1.00$\pm$3.04 &  1.00$\pm$2.07 &                                                                \\ 
 18 & 00 46 41.8 & -73 01 15 &  5.6 &    23.9 & 1.77e-03$\pm$4.42e-04 &  0.0$\pm$ 0.0 &   0.0 & 7.18E-03    & 383 & -0.26$\pm$0.10 & -0.39$\pm$0.15 & foreground star? [HFPK00]                                       \\ 
 19 & 00 46 48.1 & -73 28 07 & 10.1 &    12.4 & 3.90e-03$\pm$1.09e-03 &  5.6$\pm$ 4.2 &   2.3 & $<$1.49E-03 &    &      &      &                                                   \\ 
 20 & 00 47 08.2 & -72 43 04 &  9.6 &    11.3 & 1.55e-03$\pm$4.02e-04 &  0.0$\pm$ 0.0 &   0.0 & $<$4.83E-03 &    &      &      &                                                                \\ 
 21 & 00 47 29.2 & -73 28 23 & 10.1 &    10.4 & 2.58e-03$\pm$8.74e-04 &  0.0$\pm$ 0.0 &   0.0 &         &    &      &      &                                                   \\ 
 22 & 00 47 40.5 & -73 09 28 &  6.3 &    11.7 & 1.74e-03$\pm$5.63e-04 &  3.6$\pm$ 2.8 &   1.4 & 3.96E-02    & 419 &  1.00$\pm$0.09 &  0.27$\pm$0.04 & SNR? [KPFH99]                                                   \\ 
 23 & 00 48 19.6 & -73 31 52 &  2.3 &  5165.4 & 6.56e-02$\pm$1.99e-03 &  1.2$\pm$ 0.5 &  12.8 & 8.10E-03    & 512 & -1.00$\pm$0.24 &      & SSS RXJ0048.4-7332                                             \\ 
 24 & 00 48 42.1 & -72 46 57 &  4.8 &    17.0 & 6.73e-04$\pm$1.88e-04 &  0.0$\pm$ 0.0 &   0.0 & $<$1.83E-03 &    &      &      &                                                                \\ 
 25 & 00 48 55.6 & -73 33 37 &  3.3 &    28.7 & 1.18e-03$\pm$2.92e-04 &  0.0$\pm$ 0.0 &   0.0 & $<$2.28E-03 &    &      &      & $<$foreground star$>$ GSC 9141.7623                            \\ 
 26 & 00 48 59.0 & -72 58 13 &  5.3 &    18.9 & 1.03e-03$\pm$2.50e-04 &  0.0$\pm$ 0.0 &   0.0 & 3.18E-03    &    &      &      &  $<$foreground star$>$ \\
 27 & 00 49 08.2 & -73 21 39 &  6.4 &    10.8 & 1.23e-03$\pm$4.08e-04 &  0.0$\pm$ 0.0 &   0.0 & $<$1.35E-03 &    &      &      &                                                   \\ 
 28 & 00 49 30.8 & -73 31 09 &  3.0 &    69.4 & 2.01e-03$\pm$2.94e-04 &  2.7$\pm$ 1.7 &   7.0 & 2.39E-03    & 511 &  1.00$\pm$0.45 &  0.32$\pm$0.17 & HMXB? [HS00]                                       \\ 
 29 & 00 49 43.9 & -72 52 10 &  4.7 &    18.0 & 6.81e-04$\pm$1.89e-04 &  0.0$\pm$ 0.0 &   0.0 & $<$1.07E-02 &    &      &      &                                                                \\ 
 30 & 00 49 57.5 & -73 12 52 &  9.1 &    13.5 & 1.72e-03$\pm$4.12e-04 &  0.0$\pm$ 0.0 &   0.0 & $<$5.34E-04 &    &      &      &                                                   \\ 
 31 & 00 50 00.0 & -72 49 27 &  4.9 &    13.5 & 5.89e-04$\pm$1.79e-04 &  0.0$\pm$ 0.0 &   0.0 & $<$7.93E-03 &    &      &      &                                                                \\ 
 32 & 00 50 13.2 & -72 09 16 &  5.7 &    10.6 & 4.08e-03$\pm$1.76e-03 &  1.3$\pm$ 2.9 &   0.0 &         &    &      &      &                                                                \\ 
 33 & 00 50 19.7 & -73 35 34 &  5.7 &    11.3 & 1.28e-03$\pm$4.08e-04 &  2.2$\pm$ 2.9 &   0.2 & $<$2.26E-03 &    &      &      &                                                                \\ 
 34 & 00 50 42.9 & -73 16 08 &  1.8 &   126.2 & 4.23e-03$\pm$5.53e-04 &  1.6$\pm$ 1.4 &   1.0 & 4.52E-03    & 444 &  1.00$\pm$0.24 &  0.58$\pm$0.10 & HMXB Be star in SMC                                            \\ 
 35 & 00 50 49.3 & -72 41 53 &  4.3 &   447.9 & 1.33e-02$\pm$8.11e-04 &  0.0$\pm$ 0.0 &   0.0 &         &    &      &      & $<$foreground star$>$ SkKM 62                         \\ 
 36 & 00 50 57.2 & -73 10 06 &  3.3 &    53.4 & 2.24e-03$\pm$3.52e-04 &  0.8$\pm$ 2.8 &  0.0 & 3.48E-03    & 421 &  1.00$\pm$0.33 &  0.52$\pm$0.12 & HMXB? [HS00]                                       \\ 
 37 & 00 50 59.3 & -72 13 26 &  2.6 &   115.1 & 2.11e-02$\pm$3.71e-03 &  0.4$\pm$ 1.1 &   0.0 & $<$7.20E-03 &    &      &      & HMXB AXJ0051-722 [KP96]                                   \\ 
 38 & 00 51 02.4 & -73 21 31 &  2.7 &    50.2 & 5.49e-03$\pm$6.58e-04 &  6.2$\pm$ 2.3 &  56.6 & 1.41E-01    & 461 &  0.79$\pm$0.01 & -0.26$\pm$0.02 & SNR 0049-73.6                                                  \\ 
 39 & 00 51 20.5 & -72 16 41 &  4.6 &    10.1 & 3.53e-03$\pm$1.59e-03 &  0.0$\pm$ 0.0 &   0.0 & 1.15E-02    & 188 &  1.00$\pm$0.80 &  1.00$\pm$0.56 & XB? [KPFH99]                                                    \\ 
 40 & 00 51 30.9 & -72 25 04 &  7.3 &    11.0 & 1.11e-03$\pm$3.10e-04 &  0.0$\pm$ 0.0 &   0.0 &         &    &      &      &                                                   \\ 
 41 & 00 51 52.0 & -73 10 32 &  1.5 &   193.6 & 3.55e-03$\pm$3.83e-04 &  0.5$\pm$ 1.1 &   0.0 & 1.46E-02    & 424 &  1.00$\pm$0.08 &  0.36$\pm$0.06 & HMXB RXJ0051.9-7311 [CSM97],[HS00]                \\ 
 42 & 00 51 59.6 & -73 29 29 &  8.1 &    11.0 & 1.77e-03$\pm$5.25e-04 &  5.1$\pm$ 4.5 &   0.9 & 9.39E-03    & 496 &  1.00$\pm$0.41 &  0.35$\pm$0.25 &                                           \\ 
 43 & 00 52 07.7 & -72 25 50 & 15.9 &    10.3 & 1.19e-02$\pm$3.69e-03 &  6.3$\pm$ 9.8 &   0.1 &         &    &      &      & HMXB SMC X-3                                                   \\ 
 44 & 00 52 13.9 & -73 19 18 &  1.1 & 23375.0 & 6.79e-01$\pm$5.10e-03
&  1.1$\pm$ 0.6 &   4.5 & 2.15E-02    & 453 &  1.00$\pm$0.07 &
0.61$\pm$0.04 & HMXB RXJ0052.1-7319 [LPM99],[ISC99]               \\ 
 45 & 00 52 54.8 & -72 17 06 &  3.9 &    41.0 & 2.45e-03$\pm$4.59e-04 &  0.0$\pm$ 0.0 &   0.0 &         &    &      &      &                                                                \\ 
 46 & 00 52 55.2 & -71 58 06 &  3.0 &    83.6 & 2.37e-03$\pm$3.27e-04 &  2.9$\pm$ 1.9 &   3.9 & 1.21E-02    &  94 &  1.00$\pm$0.43 &  0.44$\pm$0.13 & HMXB RXJ0052.9-7158 [CSM97],[HS00]             \\ 
 47 & 00 53 08.7 & -72 39 56 &  7.7 &    15.2 & 2.22e-03$\pm$8.38e-04 &  0.0$\pm$ 0.0 &   0.0 & $<$4.03E-03 &    &      &      &                                                                \\ 
 48 & 00 53 24.1 & -72 27 20 &  3.1 &    39.3 & 1.23e-03$\pm$2.40e-04 &  1.3$\pm$ 1.6 &   0.2 & 3.82E-03    & 246 &  1.00$\pm$0.59 &  1.00$\pm$1.29 & HMXB? [HS00]                                       \\ 
 49 & 00 53 38.7 & -73 15 07 &  3.5 &    10.4 & 4.73e-04$\pm$1.61e-04 &  0.4$\pm$ 1.8 &  0.0 & $<$1.41E-03 &    &      &      &                                                                \\ 
 50 & 00 53 48.2 & -72 02 01 &  3.5 &    18.5 & 7.16e-04$\pm$1.92e-04 &  0.0$\pm$ 0.0 &   0.0 & 4.33E-03    & 104 &  1.00$\pm$0.57 &  1.00$\pm$1.06 &                                                                \\ 
\noalign{\smallskip}
\hline
\end{tabular}

\vspace{2mm}
Notes to column No 6: 
HRI count rate ist the output value from maximum likelihood algorithm
with the smallest positional error.

\vspace{1mm}
Notes to column No 9:
For sources which are listed in [HFPK00] PSPC count rates are taken from
the PSPC catalogue. 
Otherwise PSPC count rate is the 95.4\% (2$\sigma$) upper limit from
maximum likelihood algorithm of the pointing with maximum exposure. 

\end{table}

\clearpage

\addtocounter{table}{-1}
\begin{table}
\scriptsize
\caption[]{Continued}
\begin{tabular}{rrrrrccrrrccl}
\hline\noalign{\smallskip}
1~ & \multicolumn{1}{c}{2} & \multicolumn{1}{c}{3} & 4~ &
\multicolumn{1}{c}{5} & 6 & 7 & \multicolumn{1}{c}{8} &
\multicolumn{1}{c}{9} & \multicolumn{1}{c}{10} & 11 & 12 & ~~~13 \\
\hline\noalign{\smallskip}  
No & \multicolumn{1}{c}{RA} & \multicolumn{1}{c}{Dec} & \perr & \exil
& Count rate & \ext & \extl & Count rate & \multicolumn{1}{c}{No} & HR1 & HR2 & Remarks \\
 & & & & & & & & \multicolumn{1}{c}{PSPC} & \multicolumn{1}{c}{PSPC} & & & \\ 
 & \multicolumn{2}{c}{(J2000.0)} & [\arcsec] & & [\ct] & [\arcsec] &
& \multicolumn{1}{c}{[\ct]} & & & & \\ 
\noalign{\smallskip}\hline\noalign{\smallskip}
 51 & 00 53 56.0 & -72 26 52 &  2.4 &   395.2 & 7.76e-03$\pm$5.59e-04 &  3.1$\pm$ 1.2 &  56.3 & 9.20E-02    & 242 &  1.00$\pm$0.18 &  0.46$\pm$0.05 & HMXB? XTEJ0053-724, pulsar [CML98]                             \\ 
 52 & 00 54 12.5 & -72 17 50 &  4.5 &    24.1 & 2.88e-03$\pm$7.40e-04 &  2.5$\pm$ 2.7 &   0.4 &         &    &      &      &                                                                \\ 
 53 & 00 54 32.1 & -72 18 13 &  3.7 &    27.4 & 1.23e-03$\pm$2.56e-04 &  2.4$\pm$ 2.2 &   0.9 & 6.24E-03    & 196 &  1.00$\pm$1.37 &  1.00$\pm$0.59 &  $<$XB$>$ or $<$AGN$>$                                                              \\ 
 54 & 00 54 41.7 & -72 17 29 &  5.0 &    15.3 & 1.02e-03$\pm$2.56e-04 &  3.0$\pm$ 2.8 &   0.9 & $<$3.82E-03 &    &      &      &                                                   \\ 
 55 & 00 54 46.2 & -72 00 17 &  4.2 &    11.8 & 5.75e-04$\pm$1.80e-04 &  0.0$\pm$ 0.0 &   0.0 & $<$9.72E-03 &    &      &      &                                                                \\ 
 56 & 00 54 50.3 & -71 51 18 &  4.9 &    15.3 & 8.20e-04$\pm$2.20e-04 &  1.5$\pm$ 2.6 &   0.0 & 5.63E-03    &  76 &  1.00$\pm$1.21 &  1.00$\pm$0.98 &                                                                \\ 
 57 & 00 54 56.0 & -72 45 06 &  9.7 &    34.7 & 3.02e-03$\pm$4.84e-04 &  0.0$\pm$ 5.2 &   0.0 & 3.75E-03    & 324 &  1.00$\pm$0.71 &  0.58$\pm$0.14 & HMXB? [HS00]                                       \\ 
 58 & 00 54 56.5 & -72 26 52 &  2.6 &   127.8 & 2.82e-03$\pm$3.49e-04 &  1.5$\pm$ 1.5 &   0.5 & 3.68E-02    & 241 &  1.00$\pm$0.04 &  0.56$\pm$0.03 & HMXB XTEJ0055-724 [SCB99]          \\ 
 59 & 00 55 28.7 & -72 10 55 &  3.8 &    25.9 & 2.58e-03$\pm$6.66e-04 &  1.6$\pm$ 1.9 &   0.2 & 2.74E-02    & 157 &  1.00$\pm$0.04 &  0.36$\pm$0.04 & AGN? Radio SMC B0053-7227 [FHW98]                              \\ 
 60 & 00 55 35.1 & -72 28 36 &  4.6 &    21.9 & 1.17e-03$\pm$2.68e-04 &  0.6$\pm$ 2.8 &  0.0 & 4.47E-03    & 249 &  1.00$\pm$0.24 &  0.14$\pm$0.19 & hard [HFP00]\\
 61 & 00 55 40.4 & -72 20 29 &  5.0 &    11.4 & 9.54e-04$\pm$3.06e-04 &  0.0$\pm$ 0.0 &   0.0 & $<$1.17E-03 &    &      &      &                                                                \\ 
 62 & 00 55 43.2 & -72 09 21 &  4.7 &    11.6 & 3.47e-03$\pm$1.61e-03 &  0.0$\pm$ 0.0 &   0.0 & $<$1.27E-02 &    &      &      &                                                                \\ 
 63 & 00 55 43.9 & -73 23 43 &  9.0 &    11.1 & 1.52e-03$\pm$4.87e-04 &  0.0$\pm$ 0.0 &   0.0 &         &    &      &      &                                                                \\ 
 64 & 00 55 49.3 & -72 28 37 &  6.5 &    10.1 & 1.03e-03$\pm$3.35e-04 &  0.0$\pm$ 0.0 &   0.0 & $<$5.31E-03 &    &      &      &  $<$AGN$>$                                                 \\ 
 65 & 00 55 51.2 & -73 31 06 &  8.5 &    42.2 & 2.77e-03$\pm$4.38e-04 &  0.0$\pm$ 0.0 &   0.0 & 1.83E-02    & 508 &  1.00$\pm$0.53 &  0.36$\pm$0.07 &       \\ 
 66 & 00 55 59.5 & -72 52 41 &  7.9 &    12.0 & 5.58e-04$\pm$1.74e-04 &  0.4$\pm$ 2.1 &  0.0 & $<$7.02E-03 &    &      &      &                                                                \\ 
 67 & 00 56 00.0 & -71 52 37 &  9.7 &    10.6 & 1.40e-03$\pm$3.72e-04 &  6.2$\pm$ 5.8 &   0.4 &         &    &      &      &                                                                \\ 
 68 & 00 56 50.3 & -73 10 38 &  9.9 &    10.0 & 9.85e-04$\pm$2.90e-04 &  0.0$\pm$ 0.0 &   0.0 & 2.96E-03    & 426 &  1.00$\pm$0.38 &  0.34$\pm$0.15 & $<$foreground star$>$ GSC 9141.7719 \\
 69 & 00 56 56.4 & -73 30 21 &  9.8 &    11.8 & 1.23e-03$\pm$3.67e-04 &  2.2$\pm$ 4.1 &   0.0 &         &    &      &      &                                                                \\ 
 70 & 00 57 26.7 & -73 25 18 &  7.1 &   127.9 & 2.67e-03$\pm$3.45e-04 &  1.1$\pm$ 1.3 &   0.3 & 8.40E-03    & 476 &  1.00$\pm$0.35 &  0.66$\pm$0.08 &                                                                \\ 
 71 & 00 57 31.7 & -72 13 01 &  3.9 &   111.0 & 3.26e-03$\pm$3.40e-04 &  0.0$\pm$ 0.0 &   0.0 & 5.48E-03    & 170 &  1.00$\pm$0.13 &  0.38$\pm$0.09 & $<$XB$>$ or $<$AGN$>$   \\
 72 & 00 57 35.5 & -73 12 56 &  9.2 &    14.3 & 1.24e-03$\pm$3.04e-04 &  5.0$\pm$ 3.8 &   1.6 & 6.38E-03    & 435 &  1.00$\pm$0.84 &  0.25$\pm$0.16 & $<$XB$>$ or $<$AGN$>$\\
 73 & 00 57 48.5 & -72 02 34 &  6.4 &    22.9 & 3.08e-03$\pm$6.79e-04 &  0.0$\pm$ 0.0 &   0.0 & 3.03E-03    & 114 &  1.00$\pm$0.52 &  1.00$\pm$0.35 & HMXB? AXJ0058-72.0, pulsar [YK98a]                             \\ 
 74 & 00 57 49.3 & -72 07 55 &  3.5 &   236.9 & 4.51e-03$\pm$3.66e-04 &  1.2$\pm$ 1.5 &   0.2 & 8.82E-03    & 136 &  1.00$\pm$0.22 &  0.58$\pm$0.11 & HMXB? [HS00]                                       \\ 
 75 & 00 57 58.0 & -72 22 30 &  7.9 &    41.0 & 5.39e-03$\pm$8.32e-04 &  4.6$\pm$ 5.7 &   0.2 & $<$8.61E-03 &    &      &      &                                                                \\ 
 76 & 00 58 12.6 & -72 30 50 &  4.5 &   103.7 & 4.28e-03$\pm$4.77e-04 &  0.0$\pm$ 0.0 &   0.0 & 6.88E-04    &    &      &      & HMXB [SCC99]                                                \\ 
 77 & 00 58 16.0 & -72 18 04 &  4.5 &    69.4 & 3.00e-03$\pm$3.56e-04 &  3.7$\pm$ 3.0 &   1.3 & 7.93E-03    & 194 &  1.00$\pm$0.14 &  0.15$\pm$0.07 & SNR 0056-72.5                                                  \\ 
 78 & 00 58 20.7 & -72 16 18 &  6.3 &    10.9 & 6.78e-04$\pm$1.96e-04 &  2.3$\pm$ 3.0 &   0.3 & 2.32E-03    & 185 &  1.00$\pm$0.81 &  0.58$\pm$0.09 & $<$XB$>$ or $<$AGN$>$\\
 79 & 00 58 37.3 & -71 35 49 &  1.5 &   700.3 & 5.17e-02$\pm$3.77e-03 &  2.3$\pm$ 1.0 &  29.0 & 3.68E-01    &  47 & -0.99$\pm$0.00 & -1.00$\pm$0.26 & SSS 1E0056.8-7146                                              \\ 
 80 & 00 59 13.6 & -72 22 25 &  8.9 &    12.3 & 1.42e-03$\pm$3.37e-04 &  0.0$\pm$ 0.0 &   0.0 & 9.64E-03    &    &      &      & $<$foreground star$>$                                                               \\ 
 81 & 00 59 21.5 & -72 23 20 &  5.0 &    11.4 & 1.43e-03$\pm$5.84e-04 &  1.1$\pm$ 2.0 &   0.0 & 9.64E-03    & 218 &  1.00$\pm$0.41 &  0.52$\pm$0.06 & XB? [KPFH99]                                                    \\ 
 82 & 00 59 26.2 & -72 10 02 &  4.0 &    30.5 & 3.10e-03$\pm$3.37e-04 &  5.8$\pm$ 2.0 &  56.5 & 5.73E-02    & 148 &  0.90$\pm$0.02 & -0.12$\pm$0.03 & SNR 0057-72.2                                                  \\ 
 83 & 01 00 04.6 & -71 57 24 &  8.9 &    12.6 & 1.50e-03$\pm$3.53e-04 &  0.0$\pm$ 0.0 &   0.0 & 2.46E-03    &  91 &  1.00$\pm$1.30 &  1.00$\pm$1.93 & XB? or AGN? [KPFH99]                                            \\ 
 84 & 01 00 08.9 & -72 57 46 &  8.3 &    12.6 & 7.87e-04$\pm$2.34e-04 &  0.0$\pm$ 0.0 &   0.0 &         &    &      &      &                                                                \\ 
 85 & 01 00 12.0 & -72 20 17 &  5.3 &    10.3 & 1.35e-03$\pm$5.71e-04 &  0.0$\pm$ 0.0 &   0.0 & 1.62E-03    & 204 &  1.00$\pm$0.99 &  1.00$\pm$0.58 &  $<$XB$>$ or $<$AGN$>$\\
 86 & 01 00 13.3 & -73 07 24 &  7.2 &   234.9 & 6.94e-03$\pm$6.05e-04 &  0.0$\pm$ 0.0 &   0.0 & 2.94E-02    & 408 & -0.27$\pm$0.10 &  0.23$\pm$0.15 & foreground star? [KPFH99]                                       \\ 
 87 & 01 00 13.2 & -72 41 20 &  4.6 &    13.1 & 5.51e-04$\pm$1.70e-04 &  0.0$\pm$ 0.0 &   0.0 & 1.56E-03    & 301 &  1.00$\pm$0.52 &  1.00$\pm$0.92 &                                                                \\ 
 88 & 01 00 15.0 & -72 04 41 &  4.1 &    48.4 & 1.66e-03$\pm$2.47e-04 &  2.7$\pm$ 2.0 &   1.9 & 1.43E-03    & 123 &  1.00$\pm$0.38 &  1.00$\pm$0.57 & XB? [KPFH99]                                                    \\ 
 89 & 01 00 36.5 & -73 00 35 &  7.1 &   164.9 & 3.64e-03$\pm$4.18e-04 &  1.2$\pm$ 1.2 &   0.5 & 2.29E-02    &    &      &      &  $<$foreground star$>$                                                              \\ 
 90 & 01 00 42.2 & -72 11 33 &  3.3 &  1610.5 & 1.59e-02$\pm$6.34e-04 &  2.1$\pm$ 0.9 &  22.5 & 2.67E-02    & 162 &  1.00$\pm$0.07 &  0.33$\pm$0.04 & AGN? [KPFH99]                                                   \\ 
 91 & 01 00 49.2 & -72 03 46 &  5.5 &    11.4 & 7.20e-04$\pm$1.97e-04 &  0.0$\pm$ 0.0 &   0.0 & $<$9.15E-03 &    &      &      &                                                                \\ 
 92 & 01 00 56.2 & -72 33 54 &  4.7 &    14.7 & 7.57e-04$\pm$2.00e-04 &  2.4$\pm$ 1.9 &   1.9 & 4.16E-03    & 273 &  1.00$\pm$1.61 &  1.00$\pm$0.47 & foreground star HD 6171                                        \\ 
 93 & 01 01 02.1 & -72 06 56 &  3.7 &   181.6 & 1.15e-02$\pm$1.31e-03 &  2.3$\pm$ 1.8 &   1.7 & 5.37E-02    & 132 &  1.00$\pm$0.15 &  0.47$\pm$0.04 & HMXB RXJ0101.0-7206 [KP96],[SCB99]                        \\ 
 94 & 01 01 19.3 & -72 58 21 &  8.1 &    10.8 & 5.54e-04$\pm$1.85e-04 &  1.7$\pm$ 2.2 &   0.1 &         &    &      &      &                                                                \\ 
 95 & 01 01 19.5 & -72 11 18 &  3.6 &   235.1 & 5.61e-03$\pm$4.24e-04
&  3.4$\pm$ 1.9 &   5.9 & 1.51E-02    & 159 &  1.00$\pm$0.98 &
1.00$\pm$0.47 & HMXB? [HS00]                                       \\ 
 96 & 01 01 36.4 & -72 04 18 &  4.3 &    96.4 & 3.91e-03$\pm$4.05e-04 &  0.0$\pm$ 0.0 &   0.0 & 8.93E-03    & 121 &  1.00$\pm$0.29 &  0.18$\pm$0.09 & HMXB? [HS00]                                       \\ 
 97 & 01 01 53.3 & -72 23 35 & 10.3 &    15.2 & 1.95e-03$\pm$4.45e-04 &  0.0$\pm$ 0.0 &   0.0 & 9.98E-03    & 220 &  1.00$\pm$0.28 &  0.68$\pm$0.08 & HMXB? [HS00]                                       \\ 
 98 & 01 02 41.8 & -72 32 40 &  4.9 &    44.1 & 2.19e-03$\pm$3.48e-04 &  0.0$\pm$ 0.0 &   0.0 & 6.73E-03    & 267 &  1.00$\pm$0.33 &  0.49$\pm$0.08 & AGN? [KPFH99]                                                   \\ 
 99 & 01 02 44.1 & -72 15 22 &  4.4 &    24.6 & 1.44e-03$\pm$3.67e-04 &  0.0$\pm$ 0.0 &   0.0 & 4.05E-03    & 182 &  1.00$\pm$0.55 &  0.08$\pm$0.15 &                                                                \\ 
100 & 01 02 51.2 & -72 44 36 & 14.3 &    11.0 & 7.88e-03$\pm$2.59e-03 &  0.0$\pm$ 0.0 &   0.0 & $<$1.19E-02 &    &      &      &                                                   \\ 
\noalign{\smallskip}
\hline
\end{tabular}
\end{table}

\clearpage

\addtocounter{table}{-1}
\begin{table}
\scriptsize
\caption[]{Continued}
\begin{tabular}{rrrrrccrrrccl}
\hline\noalign{\smallskip}
1~ & \multicolumn{1}{c}{2} & \multicolumn{1}{c}{3} & 4~ &
\multicolumn{1}{c}{5} & 6 & 7 & \multicolumn{1}{c}{8} &
\multicolumn{1}{c}{9} & \multicolumn{1}{c}{10} & 11 & 12 & ~~~13 \\
\hline\noalign{\smallskip}  
No & \multicolumn{1}{c}{RA} & \multicolumn{1}{c}{Dec} & \perr & \exil
& Count rate & \ext & \extl & Count rate & \multicolumn{1}{c}{No} & HR1 & HR2 & Remarks \\
 & & & & & & & & \multicolumn{1}{c}{PSPC} & \multicolumn{1}{c}{PSPC} & & & \\ 
 & \multicolumn{2}{c}{(J2000.0)} & [\arcsec] & & [\ct] & [\arcsec] &
& \multicolumn{1}{c}{[\ct]} & & & & \\ 
\noalign{\smallskip}\hline\noalign{\smallskip}
101 & 01 03 13.3 & -72 09 17 &  3.8 &    48.4 & 2.30e-03$\pm$4.40e-04
&  1.8$\pm$ 1.7 &   1.1 & 1.54E-02    & 143 &  1.00$\pm$0.10 &
0.24$\pm$0.06 & HMXB SAXJ0103.2-7209 [HS94],[ISC98]        \\ 
102 & 01 03 14.9 & -70 50 57 &  7.4 &    43.5 & 2.84e-03$\pm$4.30e-04 &  4.2$\pm$ 2.0 &  18.5 &         &    &      &      &                                                                \\ 
103 & 01 03 26.1 & -70 53 39 &  7.2 &    80.3 & 2.54e-03$\pm$3.84e-04 &  1.7$\pm$ 1.4 &   1.8 &         &    &      &      &                                                                \\ 
104 & 01 03 28.8 & -72 47 27 &  9.7 &    40.2 & 3.88e-03$\pm$5.84e-04 &  0.0$\pm$ 0.0 &   0.0 & 1.31E-02    & 334 &  1.00$\pm$0.33 & -0.08$\pm$0.07 & AGN? [KPFH99]                                                   \\ 
105 & 01 03 37.2 & -72 01 35 &  3.6 &    53.9 & 1.72e-03$\pm$3.07e-04 &  0.6$\pm$ 1.3 &   0.0 & 1.15E+00    & 106 &  1.00$\pm$7.13 &  0.41$\pm$0.07 & HMXB? [HS00]                                       \\ 
106 & 01 03 53.9 & -72 54 41 &  7.3 &   197.6 & 6.00e-03$\pm$5.65e-04 &  0.0$\pm$ 0.0 &   0.0 & 4.40E-02    & 361 & -1.00$\pm$0.10 &      & SSS RXJ0103.8-7254 [KP96],[KPFH99]                              \\ 
107 & 01 04 00.3 & -72 02 03 &  3.3 &  9732.8 & 2.47e-01$\pm$2.75e-03 &  5.8$\pm$ 0.8 &8561.7 & 2.24E+00    & 107 &  0.92$\pm$0.00 & -0.20$\pm$0.01 & SNR 0102-72.3                                                  \\ 
108 & 01 04 35.9 & -72 21 48 &  4.0 &    21.9 & 9.63e-04$\pm$2.63e-04 &  0.0$\pm$ 0.0 &   0.0 &         &    &      &      & HMXB? [HS00]                                       \\ 
109 & 01 05 02.6 & -72 22 57 &  3.6 &    92.7 & 6.71e-03$\pm$5.16e-04 &  6.3$\pm$ 1.8 & 161.0 & 3.14E-01    & 217 &  0.66$\pm$0.01 & -0.32$\pm$0.02 & SNR 0103-72.6                                                  \\ 
110 & 01 05 09.3 & -72 11 50 &  7.8 &    13.2 & 1.58e-03$\pm$3.99e-04
&  0.0$\pm$ 0.0 &   0.0 & 6.35E-03    & 163 &  1.00$\pm$0.20 &
0.40$\pm$0.09 & HMXB AXJ0105-722 [YK98b],[FHP00]                       \\ 
111 & 01 05 13.9 & -72 33 46 &  7.5 &    11.1 & 1.02e-03$\pm$2.78e-04 &  0.0$\pm$ 0.0 &   0.0 & $<$1.69E-03 &    &      &      &                                                                \\ 
112 & 01 05 33.7 & -72 13 29 &  4.7 &    40.6 & 1.76e-03$\pm$2.99e-04 &  0.0$\pm$ 0.0 &   0.0 & 7.34E-03    & 172 &  1.00$\pm$0.37 &  0.46$\pm$0.09 & AGN? [FHP00]                                                   \\ 
113 & 01 06 14.9 & -72 05 27 &  7.4 &    38.1 & 8.49e-03$\pm$9.16e-04 & 15.5$\pm$ 5.3 &  35.5 & 3.42E-02    & 125 &  1.00$\pm$0.06 &  0.19$\pm$0.04 & SNR 0104-72.3                                                  \\ 
114 & 01 06 19.9 & -72 27 52 &  3.3 &  2101.1 & 2.35e-02$\pm$9.41e-04 &  0.0$\pm$ 0.0 &   0.0 &         &    &      &      &                                                                \\ 
115 & 01 06 28.0 & -72 22 24 &  6.4 &    11.3 & 5.00e-03$\pm$2.28e-03
&  0.0$\pm$ 0.0 &   0.0 &         &    &      &      &                                                                \\ 
116 & 01 06 57.6 & -72 24 48 &  3.5 &   136.4 & 3.08e-03$\pm$3.57e-04 &  2.3$\pm$ 1.4 &   5.7 & 1.20E-02    & 230 &  1.00$\pm$0.13 &  0.43$\pm$0.06 & hard [HFP00]\\
117 & 01 08 37.5 & -72 25 01 &  3.8 &    45.9 & 1.29e-03$\pm$2.38e-04 &  1.0$\pm$ 1.5 &   0.1 & 4.93E-03    & 232 &  1.00$\pm$0.23 &  0.39$\pm$0.10 & hard [HFP00]\\
118 & 01 17 05.1 & -73 26 37 &  0.7 & 26285.8 & 3.12e+00$\pm$5.42e-02 &  1.9$\pm$ 0.4 & 289.2 & 5.82E+00    & 482 &  0.97$\pm$0.00 &  0.39$\pm$0.00 & HMXB SMC X-1                                                   \\ 
119 & 01 17 28.9 & -73 11 44 &  4.2 &    12.0 & 2.24e-03$\pm$7.85e-04 &  2.3$\pm$ 2.4 &   0.9 & 5.51E-03    & 431 &  1.00$\pm$0.54 &  0.30$\pm$0.09 & foreground star G [SCC99]                                      \\ 
120 & 01 18 37.9 & -73 25 27 &  0.9 &  1546.7 & 2.74e-02$\pm$1.21e-03 &  2.9$\pm$ 1.1 &  40.5 & 1.11E-01    & 478 &  0.12$\pm$0.02 &  0.07$\pm$0.02 & foreground star HD 8191 [CSM97]                                \\ 
121 & 01 19 39.2 & -73 27 32 &  7.0 &    13.8 & 1.50e-03$\pm$3.90e-04 &  0.0$\pm$ 0.0 &   0.0 & 2.84E-03    & 488 & -0.19$\pm$0.15 & -1.00$\pm$1.95 &  $<$foreground star$>$                                                              \\ 
\noalign{\smallskip}
\hline
\end{tabular}
\end{table}
\label{wholecat}

\end{landscape}

\end{document}